# Orbital motion of NGC 6166 (3C 338) and its impact on the jet morphology at kiloparsec scales

A. S. R. Antas,[1]⋆ A. Caproni,[1] R. E. G. Machado,[2] T. F. Laganá,[1] G. S. Souza[1]
[1] *Núcleo de Astrofísica, Universidade Cidade de São Paulo R. Galvão Bueno 868, Liberdade, São Paulo, SP, 01506-000, Brazil*
[2] *Departamento Acadêmico de Física, Universidade Tecnológica Federal do Paraná, Av. Sete de Setembro 3165, Curitiba, PR, 80230-901, Brazil*

**ABSTRACT**
In the central region of the galaxy cluster Abell 2199 (A 2199) resides the cD galaxy NGC 6166, which spatially coincides with the 3C 338 radio source. Lobes, jets, and a more detached southern structure (similar to a jet labelled as ridge) are seen at kiloparsec-scale images of 3C 338. This unusual radio morphology has led to the proposition of different hypotheses about its physical origin in the literature. In this work, we study the feasibility of a dynamical scenario where NGC 6166 moves around the X-ray inferred centre of A 2199 from the point of view of three-dimensional hydrodynamic simulations. The physical characteristics of the intra-cluster medium in which the jet propagates are constrained to those derived from X-ray observations in the vicinity of NGC 6166. Possible orbits for the jet inlet region are derived from the estimated radial velocity of NGC 6166, while the jet parameters are constrained by parsec-scale interferometric radio observations and the estimated jet power of 3C 338 obtained from radio and X-ray data. Our results show that the hypothesis of NGC 6166 has been moving around the centre of A 2199 during the last tens of million of years is compatible with the general radio morphology of 3C 338. Furthermore, the proposed dynamic scenario for the motion of NGC 6166 may be linked to gravitational perturbations induced by the passage of a sub-cluster of galaxies hundreds of millions of years ago.

## 1 INTRODUCTION

Jets associated with active galactic nuclei (AGN) show a variety of morphologies and dynamic properties, which are determined not only by intrinsic parameters of the central engine, but also by interactions with their surroundings. Besides jet precession (e.g., Gower et al. 1982; Ruiz et al. 2013; Nawaz et al. 2016; Krause et al. 2019; Horton et al. 2020; Nandi et al. 2021), density and/or temperature gradients and relative motions between the intra-cluster medium (ICM) and the radio galaxy can produce additional distortions in the shape of jets and counter-jets (e.g., Rudnick & Owen 1976; Owen & Rudnick 1976; O'Dea & Owen 1985; O'Donoghue et al. 1993; Sakelliou & Merrifield 2000). Indeed, recent numerical simulations have further demonstrated the impact of external winds on the jet morphology, showing how colliding jets can be significantly altered by interactions with an external wind environment (e.g., Musoke et al. 2020).

Classified as a cD galaxy with AGN activity, NGC 6166 is the brightest cluster galaxy (BCG) of the galaxy cluster Abell 2199 (A 2199). Located at the central region of this cluster, it has a bright elliptical structure (labelled as component A by Minkowski 1961) surrounded by a large diffuse stellar halo at optical wavelengths (e.g. Bender et al. 2015). Three additional cluster galaxies labelled components B, C and D are found to reside close to Component A on the plane of sky (e.g., Minkowski 1961; Burbidge 1962; Lachieze-Rey et al. 1985; Lauer 1986).

The radio source 3C 338 is associated with component A of NGC 6166 (hereafter, only NGC 6166). Its parsec-scale radio maps reveal a compact core from which jet and counter-jet emanate, reaching symmetrically an approximate projected distance of 10 pc. (e.g. Feretti et al. 1993; Gentile et al. 2007; Yuan et al. 2018). Although the relatively compact and symmetrical structure seen at milliarcsecond scales, kiloparsec-scale radio maps of 3C 338 exhibit a much more complex morphology (e.g., Minkowski 1958; Edge et al. 1959; Parker & Kenderine 1967; MacDonald et al. 1968; Burns et al. 1983), as it can be seen in Figure 1. The brightest feature is the core, which coincides with the optical nucleus of NGC 6166. Symmetric jets (labelled as new jets in this work) have been ejected from the core along East-West direction, similar orientation observed at parsec-scales, ending their propagation at the location of the Eastern and Western inner lobes displayed in Figure 1.

Using observations at 1.47 and 4.87 GHz, Burns et al. (1983) estimated the spectral index, $\alpha$, along 3C 338. The core region and the western inner lobe presents respectively values of 0.5 and 0.9[1], compatible with an optically thin synchrotron emission. On the other hand, spectral indices between 1.6 and 1.9 were found in more diffuse regions associated with the lobes 1, 2, East and West, as well as in a southern structure labelled as ridge by Burns et al. (1983). The steeper values of $\alpha$ suggest that the radio lobes and the ridge are relatively older than the new jets and the core. Burns et al. (1983) also estimated the approximate age of the ridge and lobes 1 and East lobe. From arguments involving synchrotron losses and the breakdown frequency of the observed synchrotron spectrum, they estimated an age of about 9 Myr for the lobe 1, 30 Myr for the East lobe and 8 Myr for the ridge, indicating an age gradient among these structures. Using X-ray observations, Nulsen et al. (2013) estimated an age of about 7.6 Myr for the ridge, which agrees with the ridge's age inferred previously by Burns et al. (1983). Moreover, Nulsen et al. (2013) determined an age of about 5 Myr for the lobe-like structure in the end of the new jets.

---
⋆ E-mail: abraao.antas@cs.unicid.edu.br (ASRA)

[1] We have adopted in this work $S_\nu \propto \nu^{-\alpha}$, where $S_\nu$ is the flux density at a frequency $\nu$.

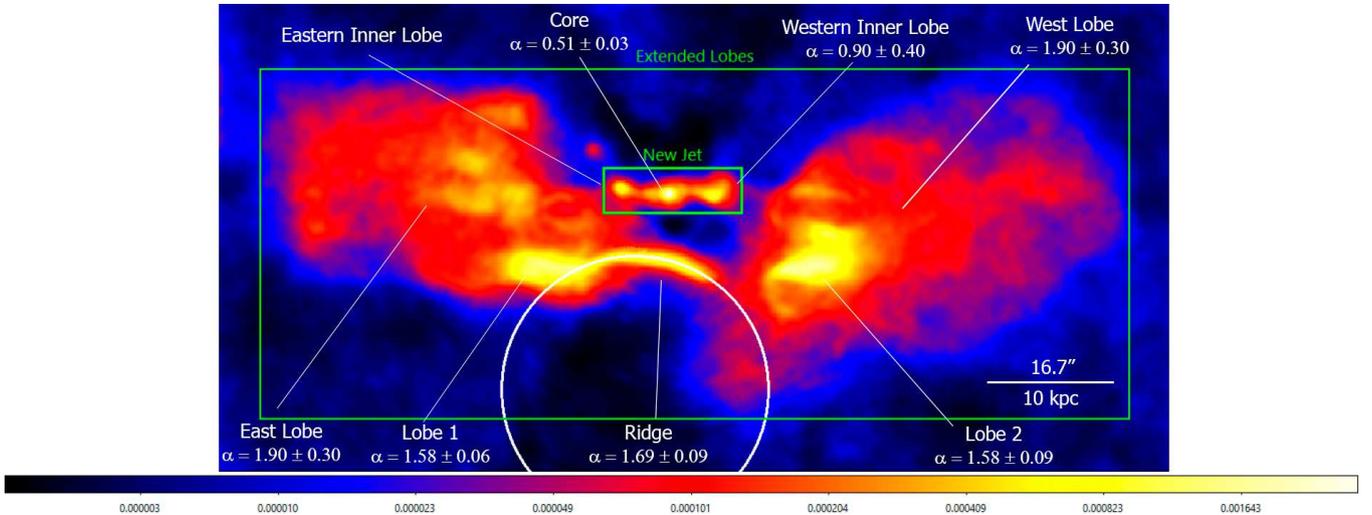

**Figure 1.** The brightness distribution of 3C 338 at 4.9 GHz (in logarithm scale and in units of Jy beam$^{-1}$ (a grey-scale version of this image is found in figure 1(a) of Ge & Owen 1994). The large and small green rectangles have dimensions of about 70 kpc × 28 kpc and 11 kpc × 4 kpc, respectively. The circular white arc has a radius of ∼ 10.8 kpc and it is used to highlight the ridge's curvature. Spectral indices for the different radio structures provided by Burns et al. (1983) are also shown in this figure.

Among the different radio structures shown in Figure 1, the most peculiar one is the ridge, a unusual jet-like feature displaced southward, where no object that could be responsible for its generation has been detected from available observations at different wavelengths. Different hypotheses about its physical origin has been proposed in the literature: orbital motion of NGC 6166 around the barycentre of the system NGC 6166–NGC 6166B, harmonic oscillations of NGC 6166 about the bottom of the gravitational potential well of the cluster, and ram pressure produced by a cooling accretion flow onto NGC 6166 (Burns et al. 1983; Nulsen et al. 2013).

Moreover, previous works have revealed strong links among the central AGN, the X-ray core and the cluster dynamics (Hudson et al. 2010; Rossetti et al. 2016). In particular, possible spatial offsets between the cD galaxy and the X-ray emission peak may be related to the ICM sloshing driven by off-axis collisions between galaxy clusters (e.g., Ascasibar & Markevitch 2006). This mechanism acting together with AGN feedback may have a significant impact on the physical characteristics of the ICM. An offset between the cD galaxy and the cluster centre suggests the occurrence of a probable motion of the galaxy through the cluster's environment. If this cD galaxy is jetted, complex radio structures can be formed during such orbital motions. It seems to be the case of NGC 6166, associated with the complex and intermittent radio source 3C 338 and presenting a projected offset of the order of 1 kpc from the X-ray centre of A 2199 (Lopes et al. 2018), as well as a radial velocity offset (in relation to the cluster mean velocity) of about +200 km s$^{-1}$ (e.g., Bender et al. 2015).

In order to study the interaction of the 3C 338's jets with the ICM, as well as testing a possible dynamic scenario where the radio source would be orbiting around the X-ray centre of A 2199 (similar to the orbital scenarios proposed by Burns et al. 1983), we performed in this work three dimensional (3D) hydrodynamics (HD) simulations of a pair of jets injected into an ambient with densities and temperatures compatible with those inferred from X-ray observations of A 2199. The jet inlet region is not static but moves according to closed orbits integrated previously under assumption of a gravitational potential well due to the dark matter (DM) in A 2199. In this work, we have used these simulations with the aim of producing a jet morphology compatible with the general radio kiloparsec-scale structures of 3C 338 detected at 1–5 GHz, as well as respecting approximately the age constraints imposed by radio and X-ray observations.

The present work is structured as follows. In section 2, we present the procedures involving the calculations of feasible orbits for NGC 6166 around the the centre of the cluster A 2199 implemented in all the HD simulations performed in this work. The numerical setup of the simulations (including the initial conditions of the ambient where the jet propagates and its injection characteristics) are presented in section 3. Simulation results (including jets with constant and time variable power) are given in section 4, while a discussion about these results and general implications are provided in section 5. Final remarks are presented in section 6. We assume throughout this work a ΛCDM cosmology with $H_0$ = 71 km s$^{-1}$ Mpc$^{-1}$, $\Omega_M$ = 0.27 and $\Omega_\Lambda$ = 0.73. The galaxy cluster A 2199 has a redshift of 0.030151 ± 0.000230 (Oegerle & Hill 2001), implying that 1.0 arcsecond = 0.599 kpc at this redshift and considering the above cosmological parameters.

## 2 ORBIT CALCULATIONS

We describe in this section the general procedures involving the determination of the NGC 6166's orbit around the centre of the galaxy cluster A 2199.

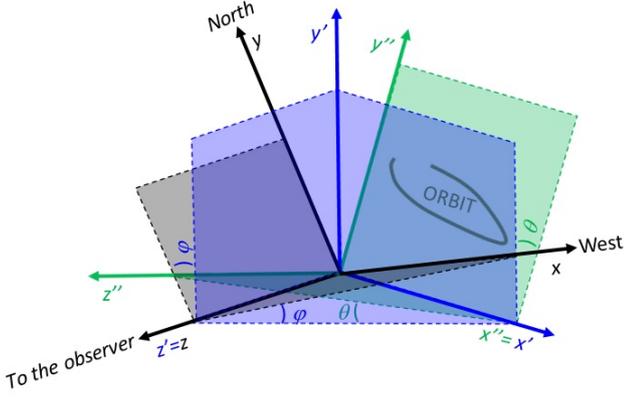

**Figure 2.** Reference frames used in the calculations of the galactic orbits. Orbital plane coincides with the $x''y''$ plane, which is rotated by an angle $\theta$ around $x''$-axis, and by an angle $\varphi$ around $z'$-axis. The axes $x$ and $y$ points respectively to West and North on the plane of the sky, while $z$-axis is aligned with the line of sight.

Let us assume that the orbital plane of NGC 6166 coincides with the $x''y''$ plane, as it is shown in Figure 2. To project this orbit on the plane of the sky ($xy$-plane in Figure 2), the original frame $O''x''y''z''$ is rotated by an angle $\theta$ around $x''$-axis and by an angle $\varphi$ around $z'$. The unit vectors $\hat{x}$, $\hat{y}$ and $\hat{z}$ (oriented respectively along the positive directions of $x$, $y$ and $z$) are related to those associated with $O''x''y''z''$ frame through

$$\begin{cases} \hat{x} = \hat{x}'' \cos\varphi + \hat{y}'' \cos\theta \sin\varphi + \hat{z}'' \sin\theta \sin\varphi, \\ \hat{y} = -\hat{x}'' \sin\varphi + \hat{y}'' \cos\theta \cos\varphi + \hat{z}'' \sin\theta \cos\varphi, \\ \hat{z} = -\hat{y}'' \sin\theta + \hat{z}'' \cos\theta. \end{cases} \quad (1)$$

We created a routine in Python[2] that uses the open package galpy[3] (Bovy 2015) to integrate numerically the orbit of NGC 6166 under a chosen DM gravitational potential, $\Phi_{tot}$. Following Hogan et al. (2017), we assumed that $\Phi_{tot}$ is composed of two different components. The first one is the Navarro-Frenk-White (NFW) gravitational potential, $\Phi_{NFW}$ (Navarro et al. 1997)

$$\Phi_{NFW}(\xi_{NFW}) = -v_{NFW}^2 \left[ \frac{\ln(1 + \xi_{NFW})}{\xi_{NFW}} \right], \quad (2)$$

where $\xi_{NFW} = r/r_{NFW}$, $r$ is the radial distance from the centre of the galaxy cluster, $r_{NFW}$ is the scale radius, $v_{NFW}^2 = 4\pi G \rho_{NFW} r_{NFW}^2$, $G$ is the gravitational constant, and $\rho_{NFW}$ is the DM mass density of the NFW profile. In this work, we fixed the virial mass at $M_{200} = 3 \times 10^{14}$ M$_\odot$, virial radius $r_{200} = 1580$ kpc and $r_{NFW} = 363.0$ kpc, which led to concentration $c = 4.35$ and $v_{NFW} = 2548.75$ km s$^{-1}$, consistent with the values estimated by Mirakhor & Walker (2020).

The second component of $\Phi_{tot}$ is a cored, isothermal DM gravitational potential, $\Phi_c$ (Binney & Tremaine 1987):

$$\Phi_c(\xi_c) = v_c^2 \left[ \frac{1}{2} \ln(1 + \xi_c^2) + \frac{\arctan \xi_c}{\xi_c} \right], \quad (3)$$

---

[2] https://www.python.org/
[3] http://github.com/jobovy/galpy

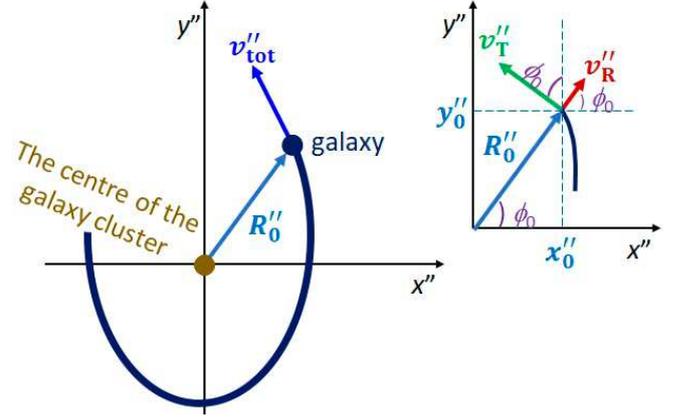

**Figure 3.** *Left panel:* Schematic representation of the orbital motion of a galaxy around the centre of its host cluster (dark blue curve) on $x''y''$ plane. The actual galactic velocity, $v''_{tot}$, and its distance from the origin, $R''_0$ (coincident with the centre of the cluster) are also shown in this figure. *Right panel:* A more detailed view emphasising the nowadays position of this galaxy ($x''_0, y''_0$), as well as its tangential and radial velocities, $V''_T$ and $V''_R$ respectively.

The angle $\phi_0$ is measured counterclockwise from the $x''$ axis.

where $\xi_c = r/r_c$, $r_c$ is the scale radius of the cored DM potential, $v_c^2 = 4\pi G \rho_c r_c^2$, and $\rho_c$ is the central DM mass density of the cored profile. Values for $v_c$ and $r_c$ adopted in the orbital scenarios in this work are listed in Table 1. We assumed $r_c = 0.1$ kpc for all simulations to keep the influential zone of the cored profile compact (its dominance in comparison with the NFW profile is limited to less than some kiloparsecs from the centre of A 2199). On the other hand, $v_c$ is chosen to keep the apocentric distance of the NGC 6166's orbit at about 6 kpc (similar to the projected distance between the X-ray centre of A 2199 and the ridge structure seen at radio wavelengths).

As initial condition for the orbit integration, it is necessary to provide the nowadays orbital-plane position of NGC 6166, ($x''_0, y''_0$), as well as the radial and tangential components of its velocity, $V''_R$ and $V''_T$, respectively. A schematic view of this initial condition is shown in Figure 3. At fixed values of $\theta$ and $\varphi$, the current galactocentric distance $R''_0$ ($= \sqrt{x''^2_0 + y''^2_0}$) and the angle $\phi_0$ (measured positively from $x''$-axis to $y''$-axis) can be determined from the relative distance between NGC 6166 and the centre of A 2199 measured on the plane of the sky. In terms of right ascension and declination offsets, $\Delta RA$ and $\Delta Dec$ respectively, $x''_0$ and $y''_0$ can be calculated from

$$\begin{cases} x''_0 = (\Delta RA \cos\varphi - \Delta Dec \sin\varphi), \\ y''_0 = \frac{(\Delta RA \sin\varphi + \Delta Dec \cos\varphi)}{\cos\theta}. \end{cases} \quad (4)$$

The quantities $\Delta RA$ and $\Delta Dec$ were estimated from the differences between the equatorial coordinates of NGC 6166 and the centre of A 2199 shown in the Table 2. The centre of A 2199 was chosen as the peak position of the X-ray surface brightness distribution obtained from CHANDRA observations of this cluster[4]. It leads to $\Delta RA =$ $-0.304$ kpc, $\Delta Dec = 0.773$ kpc, indicating a small northeastward displacement of the main body of NGC 6166 from the X-ray peak position.

---

[4] A reliable determination of the gravitational centre of a cluster of galaxies is not an easy task, relying on the complete knowledge of velocity field of the cluster galaxies, which is not achieved in practice due to the lack of direct measurements of their proper motions. An alternative method to map the mass distribution in a cluster is the gravitational lensing technique (e.g., Miralda-Escude 1991), but such measurement is not available for A 2199. As the intra-

**Table 1.** Input parameters for the orbital scenarios analysed in this work.

| Orbital Model | $v_{\text{NFW}}$ (km s$^{-1}$) | $r_{\text{NFW}}$ (kpc) | $v_c$ (km s$^{-1}$) | $r_c$ (kpc) | $V''_R$ (km s$^{-1}$) | $V''_T$ (km s$^{-1}$) | $\theta$ (deg) | $\varphi$ (deg) | Minimum $V''_{\text{tot}}$ condition? |
|---|---|---|---|---|---|---|---|---|---|
| G | 2548.75 | 363.0 | 205.50 | 0.1 | 250.75 | -418.00 | 25 | 40 | Yes |
| H | 2548.75 | 363.0 | 268.44 | 0.1 | 347.12 | -492.22 | 20 | 35 | Yes |
| I | 2548.75 | 363.0 | 268.44 | 0.1 | 347.12 | -492.22 | -20 | 215 | Yes |
| J | 2548.75 | 363.0 | 384.17 | 0.1 | 426.76 | -671.84 | -15 | 217 | Yes |
| W | 2548.75 | 363.0 | 235.97 | 0.1 | 206.37 | -500.00 | 50 | 95 | No |
| Ma | 2548.75 | 363.0 | 209.16 | 0.1 | -503.00 | -460.00 | 70 | 60 | No |
| Mb | 2548.75 | 363.0 | 209.16 | 0.1 | -1.63 | -460.00 | 70 | 101 | No |
| Mc | 2548.75 | 363.0 | 209.16 | 0.1 | -62.14 | -460.00 | 70 | 111 | No |
| Md | 2548.75 | 363.0 | 209.16 | 0.1 | -101.92 | -460.00 | 70 | 120 | No |
| O | 2548.75 | 363.0 | 235.97 | 0.1 | 19.88 | -500.00 | 70 | 101 | No |
| R | 2548.75 | 363.0 | 550.00 | 0.1 | -75.00 | -550.00 | 70 | 115 | No |
| N | 2548.75 | 363.0 | 319.50 | 0.1 | 73.63 | -600.00 | 70 | 101 | No |
| Q | 2548.75 | 363.0 | 350.00 | 0.1 | 150.00 | -650.00 | 70 | 92 | No |
| P | 2548.75 | 363.0 | 418.33 | 0.1 | 170.38 | -780.00 | 70 | 101 | No |

**Table 2.** Equatorial coordinates (J2000) assumed in this work for the centre of NGC 6166 and A 2199.

|  | Right Ascension | Declination |
|---|---|---|
| NGC 6166 [1] | 16$^h$ 28$^m$ 38$^s$.24 | 39°33′04″.23 |
| A 2199 [2] | 16$^h$ 28$^m$ 38$^s$.20 | 39°33′02″.94 |

[1] Gordon et al. 2016;
[2] X-ray emission peak estimated from CHANDRA observations.

The nowadays $x''$ and $y''$ components of the galaxy's orbital velocity, $v''_{x,\text{orb}}$ and $v''_{y,\text{orb}}$ respectively, are defined as (see Figure 3)

$$\begin{cases} v''_{x,\text{orb}} = V''_R \cos\phi_0 - V''_T \sin\phi_0 \\ v''_{y,\text{orb}} = V''_R \sin\phi_0 + V''_T \cos\phi_0 \end{cases}, \quad (5)$$

where $\cos\phi_0 = x''_0/R''_0$ and $\sin\phi_0 = y''_0/R''_0$.

Using the transformations in Equation 1, the galactic velocity on the plane of the sky is determined through

$$\begin{cases} v_{x,\text{orb}} = v''_{x,\text{orb}} \cos\varphi + v''_{y,\text{orb}} \cos\theta \sin\varphi \\ v_{y,\text{orb}} = -v''_{x,\text{orb}} \sin\varphi + v''_{y,\text{orb}} \cos\theta \cos\varphi \end{cases}, \quad (6)$$

while the velocity component of the galaxy along the line of sight, $v_{z,\text{orb}}$, is calculated from

$$v_{z,\text{orb}} = -\left(V''_R \sin\phi_0 + V''_T \cos\phi_0\right)\sin\theta. \quad (7)$$

Note that radial velocity with respect to the cluster mean velocity, $v_{\text{rad}}$ ($= -v_{z,\text{orb}}$ since $+z$ points towards the observer), is the cluster gas is usually the main baryon component in a cluster in terms of mass (e.g., David 1997), it becomes a natural probe of the gravitational potential well of the cluster. If the baryonic physics model contributes significantly to the central density profile, this could lead to offset measurements which are more effective probes of the baryonic physics than the underlying DM self-interaction properties (Roche et al. 2024) only component constrained by observations. For NGC 6166, several estimates have been made in the literature, with values ranging roughly from +156 to +378 km s$^{-1}$ (e.g., Zabludoff et al. 1990, 1993; Oegerle & Hill 2001; Coziol et al. 2009; Lauer et al. 2014; Bender et al. 2015). In this work, we adopted +206 km s$^{-1}$ for the radial velocity of NGC 6166 (Bender et al. 2015).

As the bulk velocity of NGC 6166 must be $V''_{\text{tot}} = \sqrt{V''^2_R + V''^2_T}$, we can rewrite Equation 7 as

$$\left(\frac{v_{\text{rad}}}{V''_{\text{tot}}}\right)^2 = \frac{\left[1 + \left(\frac{V''_R}{V''_T}\right) \tan\phi_0\right]^2}{\left[1 + \left(\frac{V''_R}{V''_T}\right)^2\right]\left(1 + \tan^2\phi_0\right)} \sin^2\theta. \quad (8)$$

Based on Equation 8, we divided the analyses of viable orbits for NGC 6166 in two categories: orbits integrated either from the minimum value of $V''_{\text{tot}}$ allowed for a given $\theta$ (least orbital velocity condition) or from $V''_{\text{tot}}$ higher than this minimum value (non-least orbital velocity condition).

### 2.1 Least orbital velocity condition

Fixing $\theta$ and $\phi_0$, it can be demonstrated that $v_{\text{rad}}/V''_{\text{tot}}$ in Equation 8 reaches its extremes values at

$$\left(\frac{V''_R}{V''_T}\right)_\pm = -\Upsilon \pm \sqrt{\Upsilon^2 + 1}, \quad (9)$$

where $\Upsilon = [\tan(2\phi_0)]^{-1}$.

From this relation, $V''_R \to 0$ for $\phi_0 = l\pi$, and $V''_T \to 0$ for $\phi_0 = (2l+1)\pi/2$, with $l$ being an integer in both cases. A purely radial orbit ($V''_T = 0$) implies a $\Upsilon$ that diverges. It means that $\tan\phi_0 \to 0$, leading to $x''_0 \to 0$ in Equation 4 (see also Figure 3). As a consequence, $\tan\varphi = \Delta RA/\Delta Dec$, implying to $\varphi \approx 159°$ or $339°$ for NGC 6166. Such sort of orbits were ruled out in this work since they do not cross the ridge's region in the radio maps of 3C 338, a necessary condition if the ridge is an older jet driven by the 3C 338's core millions of years ago. A glimpse of the dependency of the orbital shape in terms of $\varphi$ is provided in Appendix A for three of the orbital models explored in this work. Note that a relatively short range for $\varphi$ can produce orbits that cross the ridge at timescales similar (within a factor of 2) to the age estimate for this structure (∼ 8 Myr; Burns et al. 1983; Nulsen et al. 2013).

**Table 3.** Derived parameters for the orbital scenarios analysed in this work.

| Orbital Model | $V''_{\text{tot}}$ (km s$^{-1}$) | $R''_0$ (kpc) | $\phi_0$ (deg) | $R''_{\text{apo}}$ (kpc) | $R''_{\text{per}}$ (kpc) | $\epsilon_{\text{orb}}$ | $P_{\text{rad}}$ (Myr) | $P_{\text{az}}$ (Myr) |
|---|---|---|---|---|---|---|---|---|
| G  | 487.44 | 0.851 | 149.0 | 5.70  | 0.71 | 0.78 | 50.9 | -85.2  |
| H  | 602.30 | 0.848 | 144.8 | 6.17  | 0.66 | 0.81 | 46.5 | -76.9  |
| I  | 602.30 | 0.848 | 324.8 | 6.17  | 0.66 | 0.81 | 46.5 | -76.9  |
| J  | 795.92 | 0.839 | 327.6 | 5.97  | 0.68 | 0.80 | 35.4 | -57.2  |
| W  | 540.91 | 0.941 | 217.8 | 6.17  | 0.85 | 0.76 | 50.7 | -83.5  |
| Ma | 681.62 | 0.897 | 156.3 | 14.39 | 0.90 | 0.88 | 94.3 | -169.5 |
| Mb | 460.00 | 1.481 | 241.7 | 6.12  | 1.48 | 0.61 | 55.9 | -90.9  |
| Mc | 464.18 | 1.751 | 249.5 | 6.78  | 1.73 | 0.59 | 60.6 | -98.6  |
| Md | 471.16 | 1.970 | 254.8 | 7.43  | 1.97 | 0.58 | 64.7 | -105.6 |
| O  | 500.39 | 1.481 | 241.7 | 6.27  | 1.48 | 0.62 | 53.6 | -86.2  |
| R  | 555.09 | 2.071 | 255.0 | 4.98  | 2.07 | 0.41 | 38.0 | -57.8  |
| N  | 604.50 | 1.481 | 241.7 | 5.86  | 1.47 | 0.60 | 42.7 | -66.7  |
| Q  | 667.08 | 1.438 | 238.0 | 6.20  | 1.39 | 0.63 | 41.5 | -64.8  |
| P  | 798.39 | 1.481 | 241.7 | 6.96  | 1.43 | 0.66 | 39.8 | -62.0  |

Substituting Equation 9 into Equation 8, we obtain

$$\left(\frac{v_{\text{rad}}}{V''_{\text{tot}}}\right)^2_{\pm} = \frac{\left(1 \pm \frac{\tan \phi_0}{|\tan \phi_0|}\right)^2 \sin^2 \phi_0}{2\left[1 \pm \frac{\sin(2\phi_0)}{|\sin(2\phi_0)|}\left(2\sin^2 \phi_0 - 1\right)\right]} \sin^2 \theta. \quad (10)$$

A careful inspection of Equation 10 reveals that it can only lead to two different values: zero or $\sin^2 \theta$, depending on the value of the angle $\phi_0$. Indeed, $(v_{\text{rad}}/V''_{\text{tot}})^2 = \sin^2 \theta$ is found using the sign "+" in Equation 10 for the ranges $0 \leq \phi_0 < 90°$ and $180° \leq \phi_0 < 270°$; otherwise, the sign "-" must be used in Equation 10 for the complementary quadrants (the same applies for Equation 9).

In practice, Equation 10 can provide the least value for $V''_{\text{tot}}$ at a given $\theta$ and $v_{\text{rad}}$. The minimum $V''_{\text{tot}}$ as a function of $\theta$ considering the radial velocity inferred by Bender et al. (2015) is shown in Figure 4. For $\theta \lesssim 14°$, $V''_{\text{tot}} \gtrsim \sigma_{\text{A 2199}} = 819 \pm 32$ km s$^{-1}$ (Bender et al. 2015), where $\sigma_{\text{A 2199}}$ is the A 2199's cluster velocity dispersion. Indeed, galactic velocities higher than the dispersion velocity of their respective clusters are unusual for BCGs (e.g., Coziol et al. 2009; Ye et al. 2017; De Propris et al. 2021). Thus, a conservative lower limit of 14° for $\theta$ is assumed in this work. It is in agreement with Machado et al. (2022), who showed from $N$-body numerical simulations that the signatures of a sloshing in the ICM of A 2199 identified by Nulsen et al. (2013) may have been induced hundreds million years ago by an off-axis collision between A 2199 and a galaxy group inclined by an angle of ∼ 50° − 80° in relation to the line of sight.

### 2.2 Non-least orbital velocity condition

Even though the minimum $V''_{\text{tot}}$ condition was used to generate the orbital models G, H, I and J in Table 1, it is not an obligatory condition. Indeed, $V''_{\text{tot}}$ calculated from Equation 10 for $\theta \gtrsim 25°$ is too low ($V''_{\text{tot}} \lesssim 450$ km s$^{-1}$) to produce acceptable orbits in the case of NGC 6166: either the orbits are too compact, not allowing its radio core to cross the ridge's region seen at radio wavelengths, or the orbital periods are too long in comparison with the estimated synchrotron ages of the outer lobes of 3C 338. Thus, viable orbits for NGC 6166 are found for a relative narrow parameter range under the assumption of minimum $V''_{\text{tot}}$ shown in Figure 4.

Machado et al. (2022) found that the optimal inclination angle $\theta$ of the subcluster's orbit that drove the sloshing in A 2199 is around 70°. For a similar inclination angle $\theta$, the minimum $V''_{\text{tot}}$ assumption predicts $V''_{\text{tot}}$ between 178 and 261 km s$^{-1}$, too low to provide any acceptable orbit for NGC 6166. Thus, we relaxed the least orbital velocity condition, allowing to explore orbital models with $\theta = 70°$, labelled as M, N, O, P, Q and R, as well as an additional model with an intermediate inclination of 50° (see Table 1 for further details). In these cases, we estimated $V''_R$ from $V''_T$ and $v_{\text{rad}}$ using Equation 4 and Equation 7, pursuing a set of values for $\varphi$ that could generate acceptable orbits.

We list in Table 1 the input parameters for our routine used to integrate orbits compatible with the line of sight velocity of NGC 6166 for both least and non-least orbital velocity conditions discussed previously. In addition, some resulting parameters from orbit integration are given in Table 3. Such orbital solutions produce highly eccentric orbits ($\epsilon_{\text{orb}}$ between 0.6 and 0.8), with apocentric radius $R''_{\text{apo}}$ between 5.7 and 14.0 kpc, and pericentre radius between 0.7 and 2.1 kpc. The radial period, $P_{\text{rad}}$, ranges from 35 and 94 Myr, while the azimuthal period, $P_{\text{az}}$, is between -57 and -170 Myr (negative values because clockwise orbits in the $O''x''y''z''$ reference frame). The elapsed time assumed in the orbit integration was upper limited to 60 Myr, about a factor of 2 longer than the synchrotron lifetime of the outer lobes inferred by Burns et al. (1983).

The next step was to check whether the complex radio morphology of 3C 338 is compatible with such orbits through 3D HD simulations. The numerical setup for those simulations and their outcomes are presented in the following sections.

## 3 NUMERICAL SIMULATIONS

### 3.1 Initial setup

The classical HD simulations were conducted with the open-source code PLUTO[5] version 4.2 (Mignone et al. 2007). In HD framework, PLUTO evolves numerically in space and time the mass density, $\rho$, velocity, $v$, and thermal pressure, $P$, of a fluid through:

---
[5] http://plutocode.ph.unito.it/

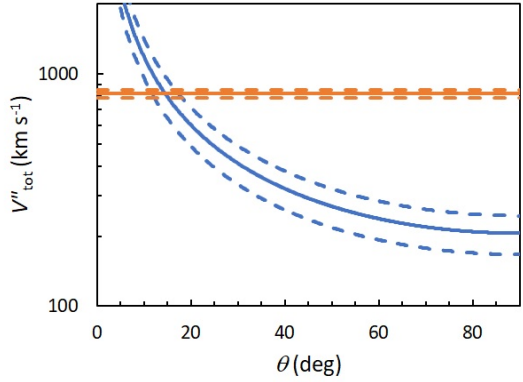

**Figure 4.** The minimum absolute velocity of NGC 6166 as a function of the inclination of its orbital plane in relation to the line of sight considering $v_{\rm rad} = +206 \pm 39$ km s$^{-1}$ (Bender et al. 2015) (blue curves, with the dashed lines showing the one-sigma uncertainties). The orange lines refer to the A 2199's cluster velocity dispersion and its one-sigma uncertainty inferred by Bender et al. (2015) (819 ± 32 km s$^{-1}$).

$$\frac{\partial \rho}{\partial t} + \nabla \cdot (\rho v) = 0, \quad (11)$$

$$\frac{\partial (\rho v)}{\partial t} + \nabla \cdot \left( \rho v v^T + P\mathbf{I} \right) = -\rho \nabla \Phi_{\rm tot}, \quad (12)$$

$$\frac{\partial E}{\partial t} + \nabla \cdot [(E + P) v] = -\rho v \cdot \nabla \Phi_{\rm tot}, \quad (13)$$

where $\mathbf{I}$ is the identity tensor of rank 3, and $E$ is the total energy density, defined as

$$E = \frac{P}{(\Gamma - 1)} + \frac{\rho |v|^2}{2}, \quad (14)$$

where $\Gamma$ is the adiabatic index of the fluid (=5/3 in this work). Note that Equation 13 is valid only if $\Phi_{\rm tot}$ is constant in time, which is indeed assumed in this work (see section 2).

The calculation of the numerical fluxes through adjacent cell interfaces were performed from the Harten, Lax, Van Leer approximate Riemann Solver for middle contact discontinuities (HLLC; Toro et al. 1994; Li 2005). The temporal evolution of $\rho$, $\rho v$, and $E$ were done via the third-order Runge–Kutta method, while a piecewise parabolic method (PPM) interpolation was applied to reconstruct the primitive variables in each time step of our simulation.

The computational domain is cubic, having a physical length of 120 kpc in each Cartesian axis. A non-uniform grid was chosen to keep the computational boundaries far from the central region of A 2199, i.e. $r < 7$ kpc, where the orbital motion of NGC 6166 takes place. A glimpse of this orbital motion is provided in Appendix B. The highest numerical resolution (0.25 and 0.5 kpc cell$^{-1}$ for simulations adopting a jet inlet radii of 1 and 2 kpc, respectively) is achieved for galactocentric distances smaller than 10 kpc, increasing by a factor of 2 in each subsequent sub volume. The numerical frontiers of the computational domain were set as selective boundary condition (Caproni et al. 2023) with a threshold speed of 0.5 km s$^{-1}$.

The simulations were run during tens of Myr (between 35 and 60 Myr, depending on the simulation), a typical timescale needed for NGC 6166 to reach its nowadays position according to the orbital scenarios discussed in section 2. Simulation outputs (spatial distributions of mass density, temperature, thermal pressure, and 3D velocity vector) were generated in time steps of 10 Myr.

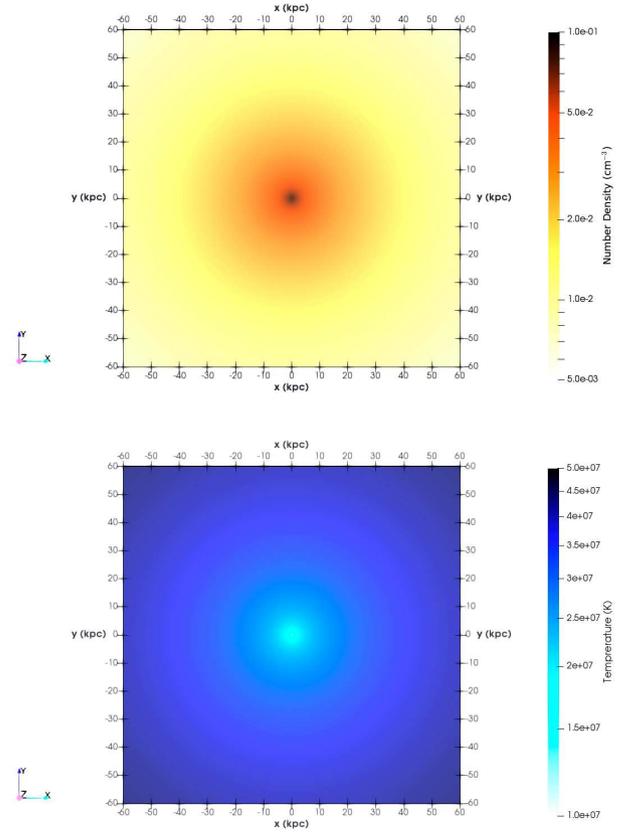

**Figure 5.** Initial spatial distribution of the number density (upper panel) and temperature (lower panel) on the plane $z = 0$ adopted in our HD simulations.

We assumed an initial static environment ($v = 0$) that has a similar number density and temperature radial profiles inferred by Johnstone et al. (2002) through X-ray observations (for further details, see figures 6 and 8 in their work). We show in Figure 5 the initial gas density and temperature distributions on the plane $z = 0$ adopted in all the simulations performed in this work. The temperature decreases by a factor of $\sim 4$ from the edges to the centre, while the number density is about ten times higher in the central region in relation to the outer parts. A non-relativistic jet is introduced in such environment, as described in subsection 3.2.

### 3.2 Jet/counter-jet injection

The injection of a pair of jets was done by selecting two bunches of adjacent cells confined within a radius $r_{\rm jet}$ where the jet number density, $n_{\rm jet}$, jet pressure, $P_{\rm jet}$, and jet velocity $v''_{\rm jet}$ are imposed. The central injection position of the jet strictly follows the orbit calculated a prior, as described in section 2. The jet is turned on at $t = 0$ and remains active until a time $t_{\rm off}$ that depends on the orbital model employed in the simulation. The jet is turned off when it is close to the ridge's position, and switched on again at a time $t_{\rm on}$ when the source reaches a location close to the current position of NGC 6166, remaining active until the end of the simulation.

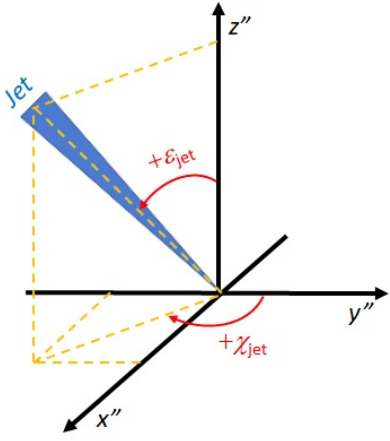

**Figure 6.** Schematic representation of the jet's orientation in relation to the $O''x''y''z''$ reference frame. The orbital motion of NGC 6166 occurs in the $x''y''$ plane.

The jet (and counter-jet) orientation in our simulations was defined through the angles $\varepsilon_{\rm jet}$, measured positively from $z''$-axis, and $\chi_{\rm jet}$, measured positively from $y''$ to $x''$ axes, as shown in Figure 6

$$\begin{cases} v''_{{\rm jet},x''} = v''_{\rm jet} \sin\varepsilon_{\rm jet} \sin\chi_{\rm jet} \\ v''_{{\rm jet},y''} = v''_{\rm jet} \sin\varepsilon_{\rm jet} \cos\chi_{\rm jet} \\ v''_{{\rm jet},z''} = v''_{\rm jet} \cos\varepsilon_{\rm jet} \end{cases}. \quad (15)$$

Applying the transformations in Equation 1 to Equation 15, we obtain the jet velocity components at $Oxyz$ reference frame

$$\begin{cases} v_{{\rm jet},x} = v''_{{\rm jet},x''}\cos\varphi + v''_{{\rm jet},y''}\cos\theta\sin\varphi + v''_{{\rm jet},z'}\sin\theta\sin\varphi \\ v_{{\rm jet},y} = -v''_{{\rm jet},x''}\sin\varphi + v''_{{\rm jet},y''}\cos\theta\cos\varphi + v''_{{\rm jet},z''}\sin\theta\cos\varphi \\ v_{{\rm jet},z} = -v''_{{\rm jet},y''}\sin\theta + v''_{{\rm jet},z''}\cos\theta \end{cases}. \quad (16)$$

On the other hand, the jet viewing angle, $\vartheta_{\rm jet}$, and the jet position angle on the plane of the sky, $\eta_{\rm jet}$, can be written in terms of $v_{{\rm jet},x}$, $v_{{\rm jet},y}$, and $v_{{\rm jet},z}$ as

$$\cos\vartheta_{\rm jet} = \frac{v_{{\rm jet},z}}{v''_{\rm jet}}, \quad (17)$$

$$\tan\eta_{\rm jet} = -\frac{v_{{\rm jet},x}}{v_{{\rm jet},y}}. \quad (18)$$

Fixing the values of $\vartheta_{\rm jet}$ and $\eta_{\rm jet}$, as well as $\theta$ and $\varphi$, equations 17 and 18 can be solved to constrain $\varepsilon_{\rm jet}$ and $\chi_{\rm jet}$. Indeed, it leads to two independent and unique solutions for $\varepsilon_{\rm jet}$ and $\chi_{\rm jet}$, which are physically interpreted as a jet (counter-jet) oriented easterly (westerly) or vice-versa.

We show in Table 4 the derived values of $\varepsilon_{\rm jet}$ and $\chi_{\rm jet}$ for each orbital model listed in Table 1, as well as for three different values of $\vartheta_{\rm jet}$ in the case of the orbital model H. Kiloparsec-scale images of 3C 338 (e.g., Burns et al. 1983; Giovannini et al. 1998; Gentile et al. 2007) show that the position angle $\eta_{\rm jet}$ of the nowadays jet/counter-jet direction is about 86° (measured from North to East), similar to the position angles of the ridge and East and West lobes inferred from the same images. For this reason, we assumed $\eta_{\rm jet} = 86°$ for all the HD simulations performed in this work. In the case of $\vartheta_{\rm jet}$, Gentile et al. (2007) estimated the quantity $\beta''_{\rm jet}\cos\vartheta_{\rm jet} = 0.14 \pm 0.04$ from the flux density ratio between the parsec-scale jet and counter-jet of 3C 338, where $\beta''_{\rm jet} = v''_{\rm jet}/c$, and $c$ is the speed of light. Using this result with the projected velocity separation between parsec-scale jet components E2 and W2 (0.346 ± 0.026c), Gentile et al. (2007) found $\beta''_{\rm jet} = 0.22 \pm 0.04$ and $\vartheta_{\rm jet} = 50° \pm 11°$. Thus, we adopted $\vartheta_{\rm jet} = 50°$ for most of the numerical studies performed in this work, except for some simulations involving orbital model H, for which two additional values for this quantity were also considered (40° and 60°, in agreement with 1$\sigma$-range for $\vartheta_{\rm jet}$ inferred by Gentile et al. 2007). Although we have assumed $v''_{\rm jet} = 0.2c$ in part of our numerical simulations (in agreement with Gentile et al. 2007), we also run simulations with lower values for $v''_{\rm jet}$. The reason for this is that $\beta''_{\rm jet} = 0.22$ was derived at parsec scales, so that jet expansion, in combination with jet propagation through the galactic ambient probably decelerate it gradually until kiloparsec distances.

**Table 4.** Angles defining jet orientation in relation to the $O''x''y''z''$ framework for different orbital models and projected jet directions (East or West) on the plane of the sky.

| Orbital Model | $\vartheta_{\rm jet}$ (deg) | $\varepsilon_{\rm jet}^{\rm East}$ (deg) | $\chi_{\rm jet}^{\rm East}$ (deg) | $\varepsilon_{\rm jet}^{\rm West}$ (deg) | $\chi_{\rm jet}^{\rm West}$ (deg) |
|---|---|---|---|---|---|
| G | 50.0 | 66.9 | 222.4 | 39.4 | 77.6 |
| H | 40.0 | 52.7 | 223.9 | 33.6 | 84.9 |
| H | 50.0 | 62.0 | 228.0 | 42.4 | 77.1 |
| H | 60.0 | 71.5 | 231.5 | 51.5 | 71.5 |
| I | 50.0 | 62.0 | 48.0 | 42.4 | 257.1 |
| J | 50.0 | 59.1 | 48.5 | 43.2 | 249.8 |
| W | 50.0 | 100.0 | 179.2 | 0.8 | 269.7 |
| M | 50.0 | 119.6 | 173.8 | 20.9 | 195.2 |
| O | 50.0 | 119.6 | 173.8 | 20.9 | 195.2 |
| R | 50.0 | 116.9 | 162.1 | 26.9 | 217.4 |
| N | 50.0 | 119.6 | 173.8 | 20.9 | 195.2 |
| Q | 50.0 | 119.7 | 182.0 | 20.1 | 174.9 |
| P | 50.0 | 119.6 | 173.8 | 20.9 | 195.2 |

## 4 SIMULATION RESULTS

In this section, we present the results from the HD numerical simulations performed in this work. Part of them assumes a jet power, $L_{\rm jet}$, constant in time (see subsection 4.1), while for the remaining ones $L_{\rm jet}$ is allowed to change with time (subsection 4.2). The jet power for a non-relativistic and non-magnetised jet can be defined as (e.g., Perucho et al. 2017):

$$L_{\rm jet} = \left(1 + \frac{u_{\rm th,jet}}{u_{\rm kin,jet}}\right)\frac{1}{2}\rho_{\rm jet}(v''_{\rm jet})^3 A_{\rm jet}, \quad (19)$$

where $\rho_{\rm jet}$ is the mass density of the jet, and $A_{\rm jet}$ is the jet inlet cross-section area. The additional terms $u_{\rm th,jet}\left(=\frac{\Gamma}{(\Gamma-1)}P\right)$ and $u_{\rm kin,jet}\left(=1/2\rho_{\rm jet}(v''_{\rm jet})^2\right)$ are, respectively, the thermal and kinetic energy densities of the jet.

All the 3D rendering visualisations of the simulations shown hereafter were obtained after rotating the computational domain by the angles $\theta$ and $\varphi$ (given in Table 1), allowing us a straightforward comparison with the radio images of 3C 338. To check whether the projected sizes of the simulated jet are compatible with those seen at 4.9-GHz image of 3C 338 (chosen as representative of the overall shape of 3C 338 in this work), as well as its main structures coincide in position with those in the real jet, we have added contours representing the 20 and 200 $\mu$Jy beam$^{-1}$ level intensities of 3C 338 at 4.9 GHz shown in Figure 1 (same data shown in figure 1 of Nulsen et al. 2013). The green rectangles in Figure 1 provide an idea of the projected dimensions of the external lobes and the restarting 4.9-GHz jet of 3C 338. The circular white arc in Figure 1 highlights the prominent bend of the ridge structure, with the curvature radius corresponding to about 18″ or $\sim$ 10.8 kpc in linear scale. We have interpreted this bend as a consequence of the motion of the host galaxy of 3C 338 through the ICM of A 2199 (see section 5 for further details).

**Table 5.** General parameters for the HD simulations with time-independent jet power performed in this work.

| Simulation ID | $\Delta t_{\rm simul}$ (Myr) | $t_{\rm off}$ (Myr) | $t_{\rm on}$ (Myr) | $n_{\rm jet}$ ($10^{-4}$ cm$^{-3}$) | $P_{\rm jet}$ ($10^{-12}$ erg cm$^{-3}$) | $v''_{\rm jet}$ ($c$) | $u_{\rm kin,jet}/u_{\rm th,jet}$ | $r_{\rm jet}$ (kpc) | $L_{\rm jet}$ ($10^{43}$ erg s$^{-1}$) |
|---|---|---|---|---|---|---|---|---|---|
| GEcLt60n0.2v0.2rj2.0 | 60 | 45 | 57 | 0.2 | 0.28 | 0.20 | 861 | 2.0 | 42.9 |
| GWcLt60n0.2v0.2rj2.0 | 60 | 45 | 57 | 0.2 | 0.28 | 0.20 | 861 | 2.0 | 42.9 |
| HEcLt60n0.2v0.2rj2.0 | 60 | 45 | 57 | 0.2 | 0.28 | 0.20 | 861 | 2.0 | 42.9 |
| HWcLt60n0.2v0.2rj2.0 | 60 | 45 | 57 | 0.2 | 0.28 | 0.20 | 861 | 2.0 | 42.9 |
| IEcLt60n0.2v0.2rj2.0 | 60 | 45 | 57 | 0.2 | 0.28 | 0.20 | 861 | 2.0 | 42.9 |
| IWcLt60n0.2v0.2rj2.0 | 60 | 45 | 57 | 0.2 | 0.28 | 0.20 | 861 | 2.0 | 42.9 |
| JEcLt50n0.2v0.2rj2.0 | 50 | 41 | 47 | 0.2 | 0.28 | 0.20 | 861 | 2.0 | 42.9 |
| JWcLt50n0.2v0.2rj2.0 | 50 | 41 | 47 | 0.2 | 0.28 | 0.20 | 861 | 2.0 | 42.9 |
| JEcLt50n12.8v0.05rj2.0 | 50 | 41 | 47 | 12.8 | 17.8 | 0.05 | 54 | 2.0 | 43.6 |
| JEcLt50n1.6v0.1rj2.0 | 50 | 41 | 47 | 1.6 | 2.22 | 0.1 | 215 | 2.0 | 43.0 |
| WEcLt60n5.0v0.1rj2.0 | 60 | 48 | 57 | 5.0 | 6.94 | 0.1 | 215 | 2.0 | 134.4 |
| MaEcLt60n5.0v0.1rj2.0 | 60 | 48 | 57 | 5.0 | 6.94 | 0.1 | 215 | 2.0 | 134.4 |
| MbEcLt60n5.0v0.1rj2.0 | 60 | 48 | 57 | 5.0 | 6.94 | 0.1 | 215 | 2.0 | 134.4 |
| McEcLt60n5.0v0.1rj2.0 | 60 | 48 | 57 | 5.0 | 6.94 | 0.1 | 215 | 2.0 | 134.4 |
| MdEcLt60n5.0v0.1rj2.0 | 60 | 48 | 57 | 5.0 | 6.94 | 0.1 | 215 | 2.0 | 134.4 |
| MdEcLt60n8.0v0.1rj1.0 | 60 | 48 | 57 | 8.0 | 11.10 | 0.1 | 215 | 1.0 | 53.8 |
| OEcLt60n5.0v0.1rj2.0 | 60 | 48 | 57 | 5.0 | 6.94 | 0.1 | 215 | 2.0 | 134.4 |
| NEcLt60n5.0v0.1rj2.0 | 60 | 48 | 57 | 5.0 | 6.94 | 0.1 | 215 | 2.0 | 134.4 |
| QEcLt40n13.0v0.1rj1.0 | 40 | 34 | 39 | 13.0 | 132.63 | 0.1 | 30 | 1.0 | 95.3 |
| PEcLt60n5.0v0.1rj2.0 | 60 | 52 | 57 | 5.0 | 6.94 | 0.1 | 215 | 2.0 | 134.4 |

### 4.1 HD simulations with time constant jet power

The HD simulations for which the jet inlet power was kept fixed in time are identified by 'cL' in their labels. The main characteristics of these simulations are presented in Table 5. The labels listed in its first column follow the general structure 'MDpLtNNnN.NvN.NrjN.N' to facilitate the identification of a given simulation in the text, where 'M' represents the orbital model used to locate instantaneously the jet inlet region (see section 2), 'D' the pointing direction of the jet on the plane of the sky ('E' for East and 'W' for West),'p' indicates whether jet power is constant ('c') or time-varying ('v') along the simulation, 'tNN' provides the total integration time employed in units of Myr, 'nN.N' is the jet number density in units of $10^{-4}$ cm$^{-3}$, 'vN.N' refers to the jet inlet speed in units of $c$, and 'rjN.N' is the jet inlet radius in units of kpc.

We show in Figure 7 the 3D iso-contours of number density from eight simulations using four different orbital models (G, H, I and J) and considering the two possibilities of jet orientation on the plane of the sky (East and West directions). The colour hues represent the synchrotron emissivity, $j_\nu^{\rm syn}$, calculated from equipartition arguments concerning energy densities of the jet particles and the magnetic field (e.g., Mioduszewski et al. 1997):

$$j_\nu^{\rm syn} \propto \delta^{(\alpha+2)} \left(\frac{P}{\Gamma-1}\right)^{(\alpha+3)/2} \nu^{-\alpha}, \qquad (20)$$

where $\nu$ is the frequency measured in the observer's reference frame, $\alpha$ is the spectral index ($S_\nu \propto \nu^{-\alpha}$, where $S_\nu$ is the flux density), and $\delta$ is the relativistic Doppler boosting factor[6], defined as:

$$\delta = \frac{\sqrt{1-\beta^2}}{\left(1-\frac{v}{c}\cdot\hat{z}\right)}, \qquad (21)$$

where $\beta = |v|/c$.

All simulations presented in Figure 7 have their brightest region coinciding with the nowadays position of the active core of NGC 6166 and its younger jet, which is indeed seen in the 4.9-GHz map of 3C 338. Besides, the simulated outer lobes tend to curve towards North direction, with the highest bends being observed in the simulations using the orbital model J (JEcLt50n0.2v0.2rj2.0 and JWcLt50n0.2v0.2rj2.0). As the orbital speeds increase from model G to J, the relative motion between the AGN and the ICM also enhances, allowing that ram-pressure (RP) forces deflect the jets from its original direction. As jet curvature induced by RP depends on the squared ratio between galaxy and jet velocities (e.g, Gunn & Gott 1972; Begelman et al. 1979; Vallee et al. 1981; Baan & McKee 1985), it is expected (fixing the jet number density) that RP to be more gradually important from G to J models, as it is indeed noted in Figure 7.

The E-W lengths of the simulated jets in Figure 7 also exhibit a general trend: simulations with a jet pointing West (right column panels) are slightly longer than those with a jet oriented easterly (left column panels), whatever the orbital model adopted in the simulation.

---

[6] Although all the simulated jets are non-relativistic, implying $\delta \sim 1$, we maintained the term $\delta^{(\alpha+2)}$ in Equation 20, as well as in the calculations of $j_\nu^{\rm syn}$ in this work.

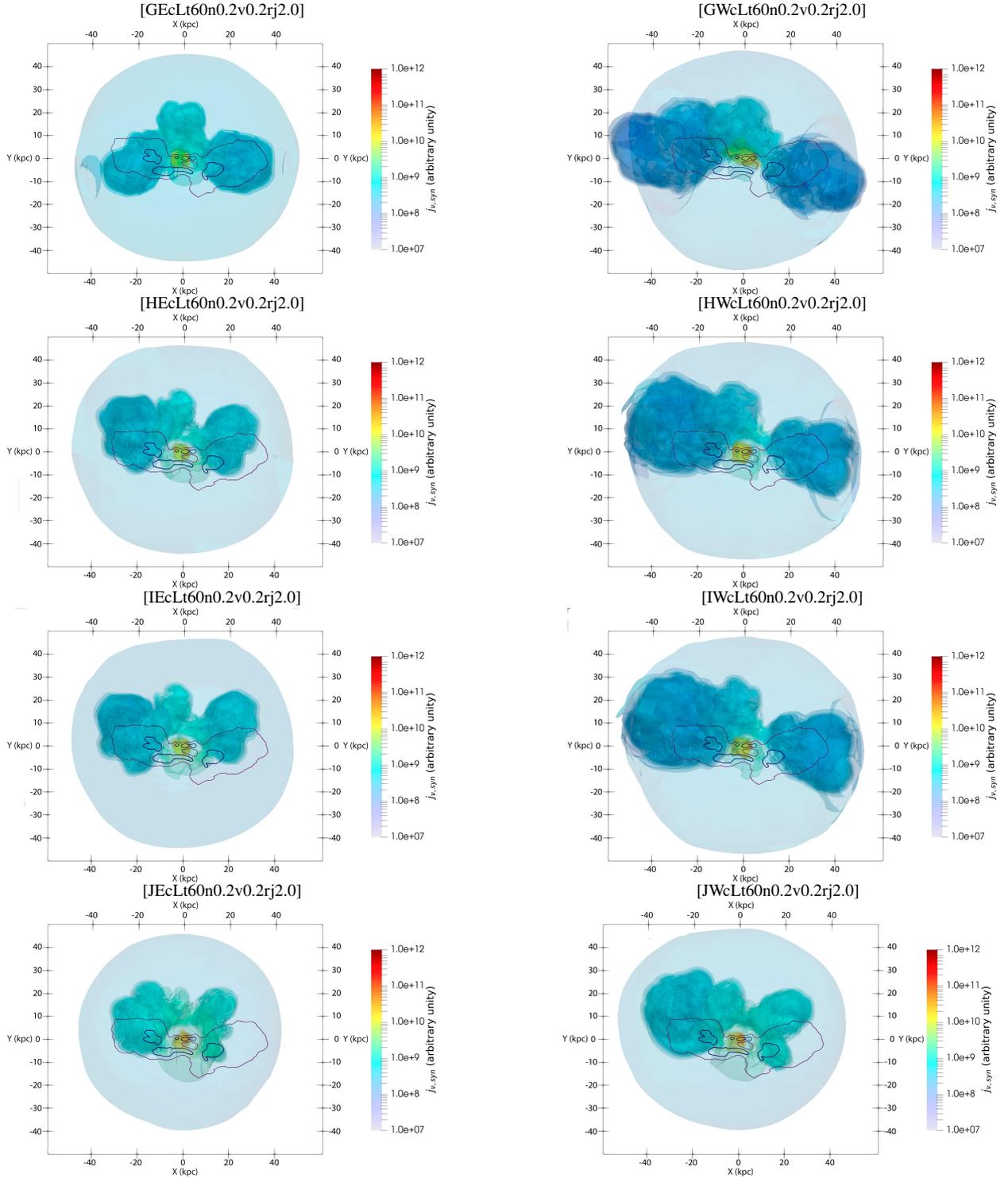

**Figure 7.** Three-dimensional iso-contours (in logarithm scale) of the number density comparing the last output of simulations with the jet pointing towards East and West directions (left and right columns, respectively). The coloured scale represents the synchrotron emissivity (in arbitrary units) using $\alpha = 1.4$ in Equation 20. Purple and blue contours refer, respectively, to 20 and 200 $\mu$Jy beam$^{-1}$ level intensities of 3C 338 at 4.9 GHz shown in Figure 1.

It comes from the differences between the values of $\varepsilon_{\rm jet}$ in the East and West jet scenarios (see Table 4); for a particular orbital model, $\varepsilon_{\rm jet}$ is always larger for East-jet solutions, implying that the jet is closer to the orbital plane and more susceptible to strong interactions with the surrounding ambient due to the galactic motion around the A 2199's centre (e.g., Baan & McKee 1985; see Appendix C for further discussion). Even though the arc-second-scale radio maps of 3C 338 cannot bring a reliable indicative of which jet (East or West) is approaching to us, the total length and flux density of the parsec-scale jet of 3C 338 suggests that the East jet is pointing to the observer (Gentile et al. 2007). Thus, we will hereafter assume that the jet (counter-jet) is oriented easterly (westerly).

We show in Figure 8 the influence of $\vartheta_{\rm jet}$ on the simulated jets after 60 Myr of evolution. We have assumed three different values for $\vartheta_{\rm jet}$ (40°, 50° and 60°) in the framework of the model HEcLt60n0.2v0.2rj2.0 (see Table 5). The increase of $\vartheta_{\rm jet}$ made the projected jet structures more compact, with the outer lobes closest to each other. The reason for that is the increase of $\varepsilon_{\rm jet}$ with the increment of $\vartheta_{\rm jet}$ (see Table 4), leading the jet closer to the orbital plane (as in the case of the differences in the elongation between West and East oriented jets shown in Figure 7; see Appendix C for further discussion).

We also analysed the influence of the specific momentum of the jet ($\rho_{\rm jet} v''_{\rm jet}$) on the overall simulated structures. From JEcLt50n0.2v0.2rj2.0, we created two additional simulations, JEcLt50n1.6v0.1rj2.0 and JEcLt50n12.8v0.05rj2.0, for which $n_{\rm jet}$ and $v''_{\rm jet}$ were varied but keeping $L_{\rm jet}$ roughly the same (see Table 5). The results after 60 Myr of evolution can be seen in Figure 9; from the top to the bottom panels $v''_{\rm jet}$ ($n_{\rm jet}$) decreases (increases) by a factor of two (eight), resulting in an increase of the specific momentum of the jet by factors of four. However, the total E-W length of the jet structure did not increase by such amount but only by a factor of $\sim 1.6$ from JEcLt50n0.2v0.2rj2.0 to JEcLt50n1.6v0.1rj2.0, and a factor of $\sim 1.3$ from JEcLt50n1.6v0.1rj2.0 to JEcLt50n12.8v0.05rj2.0. These values agree with the predicted bow-shock speed, $v_{\rm bs}$, under the highly supersonic flow condition (e.g., de Gouveia dal Pino & Benz 1993):

$$v_{\rm bs} \approx \frac{v''_{\rm jet}}{1 + \left(r_{\rm h}/r_{\rm jet}\right)\left(\rho_{\rm amb}/\rho_{\rm jet}\right)^{1/2}}, \quad (22)$$

where $r_{\rm h}$ is the bow-shock (jet head) radius and $\rho_{\rm amb}$ is the mass density of the pre-shocked ambient.

Considering a bow-shock radius similar to the jet radius ($1 \lesssim r_{\rm h}/r_{\rm jet} \lesssim 5$) and the ambient densities crossed by the jet inlet region in the orbital model J, Equation 22 provides values between 1.3 and 1.4 for the ratio between the East-West length of the jet obtained in the simulations JEcLt50n0.2v0.2rj2.0 and JEcLt50n1.6v0.1rj2.0. For the simulations JEcLt50n1.6v0.1rj2.0 and JEcLt50n12.8v0.05rj2.0, analytical estimates are roughly between 1.1 and 1.3.

We have assumed $\alpha = 1.4$ in the calculations of the pseudo synchrotron emissivity (Equation 20) shown in Figure 7, Figure 8 and Figure 9. This value corresponds to a simple arithmetic mean among the spectral index of the different regions of 3C 338 derived by Burns et al. (1983), displayed also in Figure 1. While the kiloparsec-scale core has an spectral index of $\sim 0.5$ between 6 and 20 cm, the East and West lobes exhibit a steeper index, $\sim 1.9$ (see Figure 1 in Burns et al. 1983). To show the impact of the $\alpha$ parameter on $j_\nu^{\rm syn}$, we present in Figure 10 additional 3D renderings of the model HEcLt60n0.2v0.2rj2.0 for $\alpha = 0.5, 0.9$ and $1.9$. As $j_\nu^{\rm syn} \propto P^{(\alpha+3)/2}$, high values of $\alpha$ produces higher values of $j_\nu^{\rm syn}$ for a fixed value of $P$. Even though realistic synchrotron images from our simulations must be obtained from radiative transfer calculations, as well as from the inclusion of magnetic fields and a spectral-index position-dependent distribution, Figure 10 is enough to provide a qualitative indicative of the $\alpha$'s influence on the overall shape of the simulated structures perceived by hypothetical observers in the optically-thin regime.

The simulations presented until now involves the orbital scenarios G, H, I and J, all of them representing orbits close to the plane of sky ($|\theta| \sim 15° - 25°$). However, face-on orbits are not the only possibility in the case of NGC 6166 if the minimum $V''_{\rm tot}$ condition is relaxed. Indeed, Machado et al. (2022) analysed the scenario where the cause of a sloshing feature seen at X-rays in the central region of A 2199 was due to an off-axis collision between galaxy clusters. Their $N$−body numerical simulations pointed out that a collision between A 2199 and a galaxy group ($M_{200} \sim 10^{13}$ M$_\odot$) occurring nearly edge-on ($|\theta| \sim 70°$) is able to generate structures that are consistent with the observations. Motivated by this, we created new orbital models with higher values of $\theta$, labelled W, M, O, N, P and Q (see Table 1, Table 3 and Table 4), which are more compatible with the results found by Machado et al. (2022).

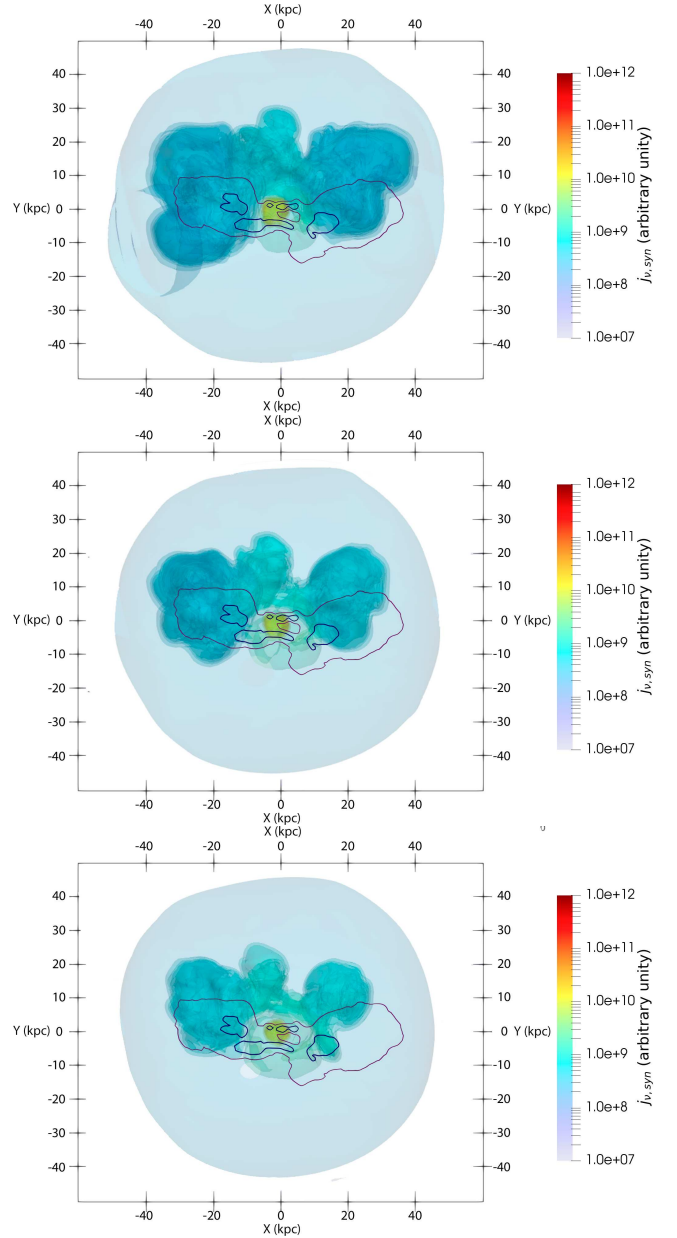

**Figure 8.** The influence of $\vartheta_{\rm jet}$ on the jet's morphology after 60 Myr of evolution and considering model HEcLt60n0.2v0.2rj2.0 (see Table 5). Top, middle and bottom panels refer respectively to $\vartheta_{\rm jet} = 40°, 50°$ and $60°$. Purple and blue contours refer, respectively, to 20 and 200 $\mu$Jy beam$^{-1}$ level intensities of 3C 338 at 4.9 GHz shown in Figure 1.

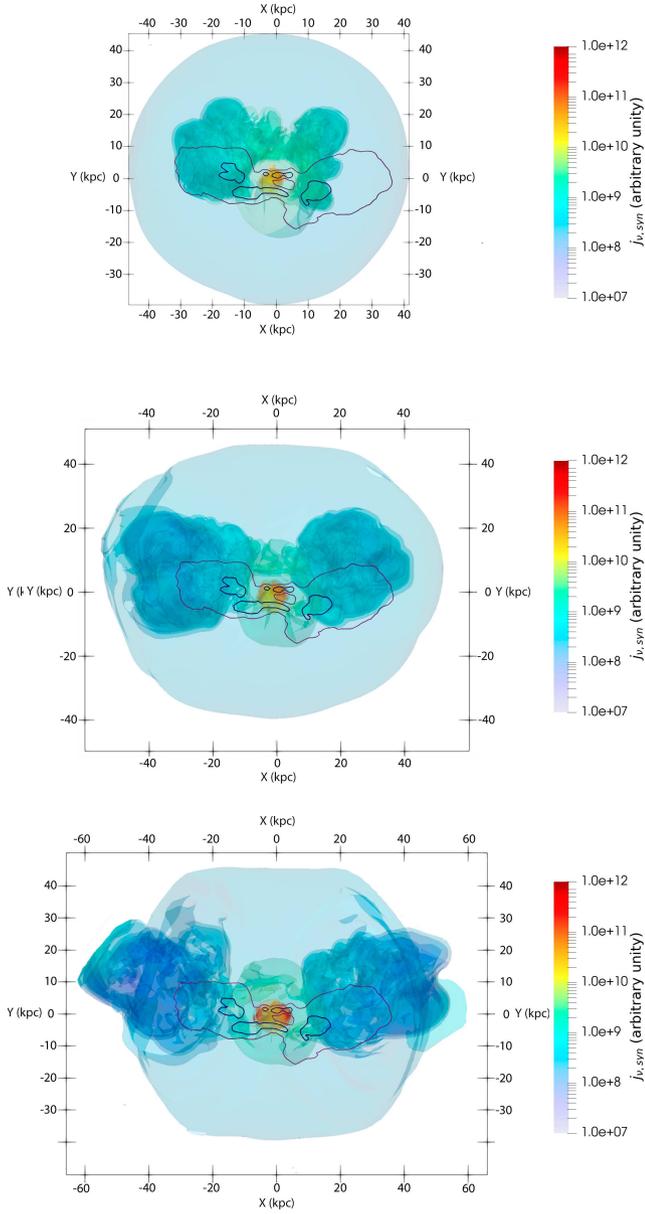

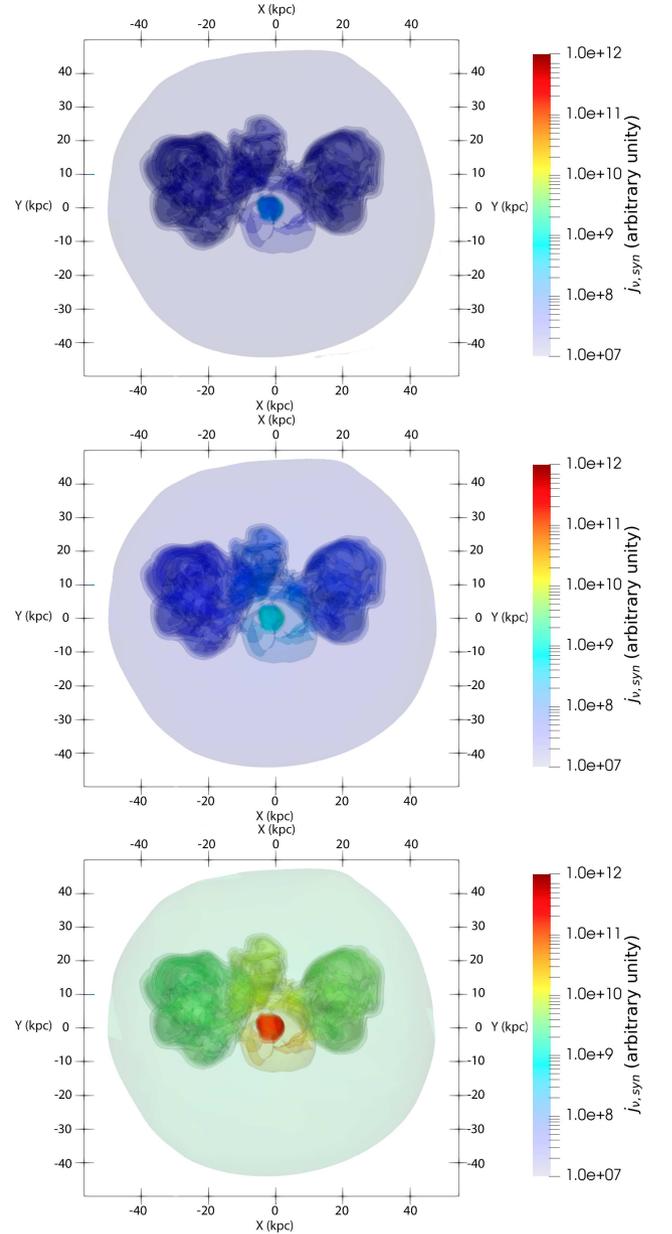

**Figure 9.** The influence of the specific momentum of the jet on the jet's morphology after 50 Myr of evolution for the orbital model J and maintaining $L_{\rm jet}$ approximately fixed. Top, middle and bottom panels refer respectively to $v''_{\rm jet} = 0.2c$ and $n_{\rm jet} = 2 \times 10^{-4}$ cm$^{-3}$ (model JEcLt50n0.2v0.2rj2.0), $v''_{\rm jet} = 0.1c$ and $n_{\rm jet} = 1.6 \times 10^{-3}$ cm$^{-3}$ (model JEcLt50n1.6v0.1rj2.0) and $v''_{\rm jet} = 0.05c$ and $n_{\rm jet} = 0.0128$ cm$^{-3}$ (model JEcLt50n12.8v0.05rj2.0). The coloured scale represents the synchrotron emissivity (in arbitrary units) using $\alpha = 1.4$ in Equation 20. Purple and blue contours refer, respectively, to 20 and 200 $\mu$Jy beam$^{-1}$ level intensities of 3C 338 at 4.9 GHz shown in Figure 1.

**Figure 10.** The influence of the spectral index on the on the synchrotron emissivity after 60 Myr of evolution for the model HEcLt60n0.2v0.2rj2.0. Top, middle and bottom panels refer respectively to $\alpha = 0.5$, 0.9 and 1.9.

Examples of simulations adopting orbital models M, O, N and P are shown in Figure 11, in which a jet with $5 \times 10^{-4}$ cm$^{-3}$ and injection velocity of $0.1c$ were assumed (see Table 5). Despite the similar synchrotron emissivities after 60 Myr of evolution, the overall structure of the jets exhibits two main trends considering the sequence MbEcLt60n5.0v0.1rj2.0, OEcLt60n5.0v0.1rj2.0, NEcLt60n5.0v0.1rj2.0 and PEcLt60n5.0v0.1rj2.0: a decrease of the jet length along E-W direction and an increase of the bend in the jet towards North. This is a consequence of the increase of the orbital velocity related to the models M, O, N and P, as shown in Table 3.

We show in Figure 12 the iso-contours of the number density coloured by synchrotron emissivity after 60 Myr of jet evolution considering the orbital model W, for which the normal of the orbital plane of the jet inlet region is inclined by 50° to the line of sight. This simulation provides an intermediate case between almost face-on orbits (models G, H, I and J) and edge-on ones (models M, O, N and P). The usage of the same jet power adopted in the previous models M, O, N and P produced jets that propagated larger distances in comparison with the jet mapped at 4.9 GHz (about 10 kpc in East-West direction), similar to the case of MbEcLt60n5.0v0.1rj2.0 shown in Figure 11.

All the simulations presented until now (Figures 7 – 12) shows a persistent structure right above the New Jet (restarting jet) that is not seen in the GHz-radio maps of 3C 338. Even though the adopted representation of the simulated jets shown in those figures is only an approximation of what should be a real radio map of the source, the comparable prominence of this structure in relation to the extended E-W lobes led us to explore some alternative to diminish its importance in the synthetic images of the simulated jets. The previous simulations tested different values for $\theta$, $\vartheta_{\rm jet}$, $\alpha$ and $v''_{\rm jet}$, but not the influence of the rotation angle $\varphi$ on the morphology of the jet for a fixed orbital motion and a specific momentum of the jet. We present in Figure 13 the results of four simulations with different values for $\varphi$ using as reference simulation MbEcLt60n5.0v0.1rj2.0 (top right panel in Figure 13). As expected, the simulation MaEcLt60n5.0v0.1rj2.0 with $\varphi = 60°$ produced a jet systematically displaced northwards in relation to the location of 3C 338 since its orbit did not cross the southern structures (e.g., ridge) during the last 60 Myr (see also the top right panel in Figure A3). Comparing the reference model MbEcLt60n5.0v0.1rj2.0 ($\varphi = 101°$) with those generated from McEcLt60n5.0v0.1rj2.0 ($\varphi = 111°$) and MdEcLt60n5.0v0.1rj2.0 ($\varphi = 120°$), we can realise that the spurious structure above the actual jet decreases its importance as $\varphi$ grows, disappearing completely when $\varphi = 120°$ is adopted. It indicates that $\varphi$ regulates the formation of this structure, favouring orbits that have their apocentric radius oriented easterly in relation to the brightest region in the ridge seen at 4.9 GHz.

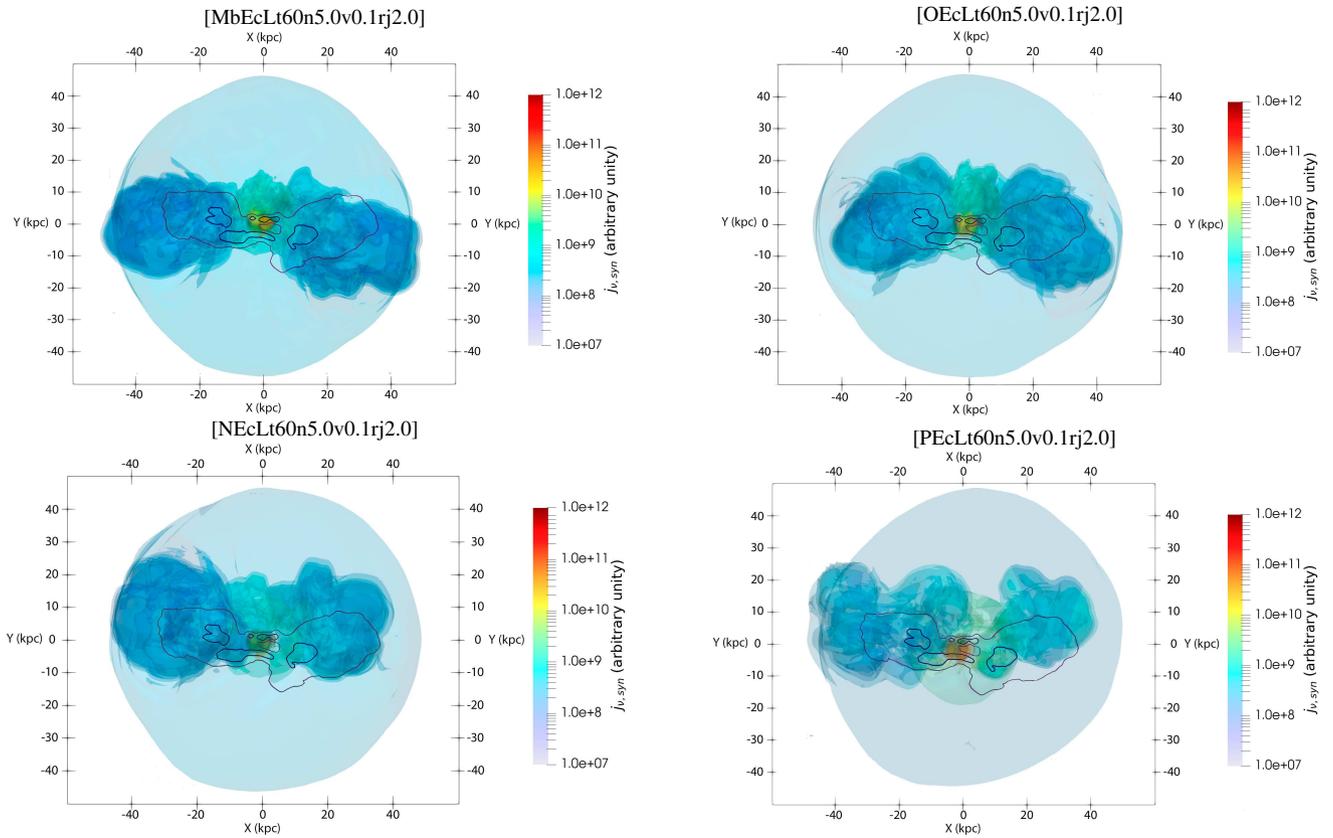

**Figure 11.** Three-dimensional iso-contours (in logarithm scale) of the number density comparing the last output of simulations with orbital planes inclined by 70° to the line of sight (models M, O, N and P) and with the jet pointing towards East. The coloured scale represents the synchrotron emissivity (in arbitrary units) using $\alpha = 1.4$ in Equation 20. Purple and blue contours refer, respectively, to 20 and 200 $\mu$Jy beam$^{-1}$ level intensities of 3C 338 at 4.9 GHz shown in Figure 1.

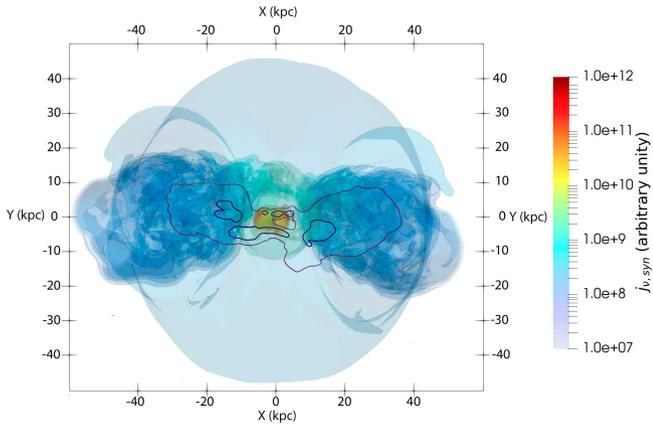

**Figure 12.** Three-dimensional iso-contours (in logarithm scale) of the number density comparing the last output of simulations with orbital plane inclined by 50° to the line of sight (model W) and with the jet pointing towards East. The coloured scale represents the synchrotron emissivity (in arbitrary units) using $\alpha = 1.4$ in Equation 20. Purple and blue contours refer, respectively, to 20 and 200 $\mu$Jy beam$^{-1}$ level intensities of 3C 338 at 4.9 GHz shown in Figure 1.

The simulations presented in Figures 7 – 13 considered a jet inlet region with a radius of 2 kpc, enough to explore the overall dependence of the jet and/or orbital parameters on the jet evolution. At this point, it is worth to verify the influence of the numerical resolution in our results. With this aim, we decreased by a factor of two the radius of the jet inlet region, keeping the same number of numerical cells per jet radius. We show in Figure 14 the last output of the simulation MdEcLt60n8.0v0.1rj1.0 using the orbital model Md, $r_{\text{jet}} = 1$ kpc, $\beta_{\text{jet}} = 0.1c$ and $n_{\text{jet}} = 8 \times 10^{-4}$ cm$^{-3}$. The overall shape of the jet is similar to that seen in the bottom right panel in Figure 13, with the advantage of the jet/counter-jet length are more compatible with the size of the outer lobes detected in 3C 338 at 4.9 GHz than in the case of simulation MdEcLt60n5.0v0.1rj2.0. Besides, the jet power in MdEcLt60n8.0v0.1rj1.0 agrees with that derived by Gentile et al. (2007) to produce the western outer lobe, in contrast with the jet power assumed in MdEcLt60n5.0v0.1rj2.0, a factor of $\sim 3$ higher than the value inferred from those observations. Hereafter, only results from simulations with $r_{\text{jet}} = 1$ kpc will be presented in the next sections.

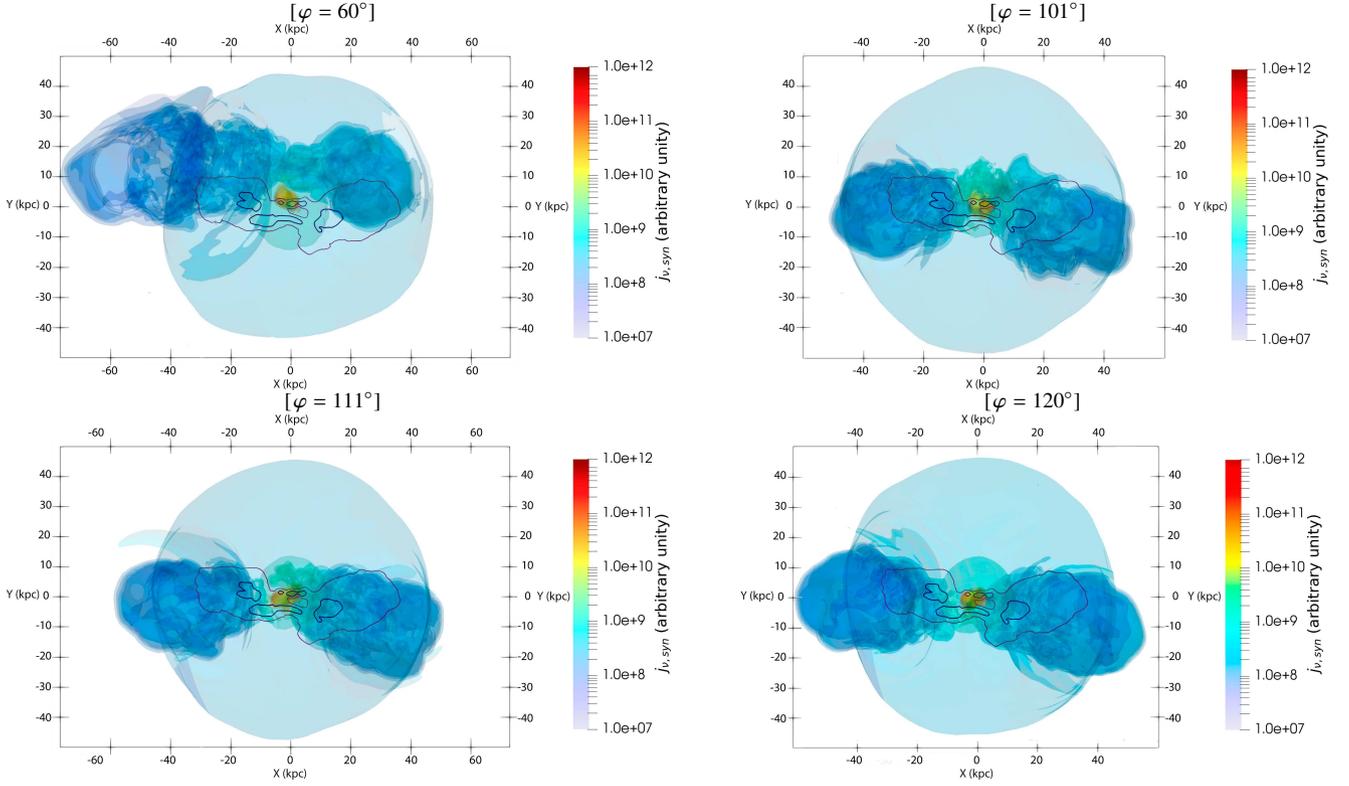

**Figure 13.** Influence of the angle $\varphi$ on the jet morphology after 60 Myr of evolution. All simulations adopt the setup used in the simulation MbEcLt60n5.0v0.1rj2.0 (top left panel in Figure 11 and shown again on the top right panel – $\varphi = 101°$ – for helping straightforward comparisons among simulations). Plots show the three-dimensional iso-contours (in logarithm scale) of the number density coloured by the synchrotron emissivity (in arbitrary units) using $\alpha = 1.4$ in Equation 20. Purple and blue contours refer, respectively, to 20 and 200 $\mu$Jy beam$^{-1}$ level intensities of 3C 338 at 4.9 GHz shown in Figure 1.

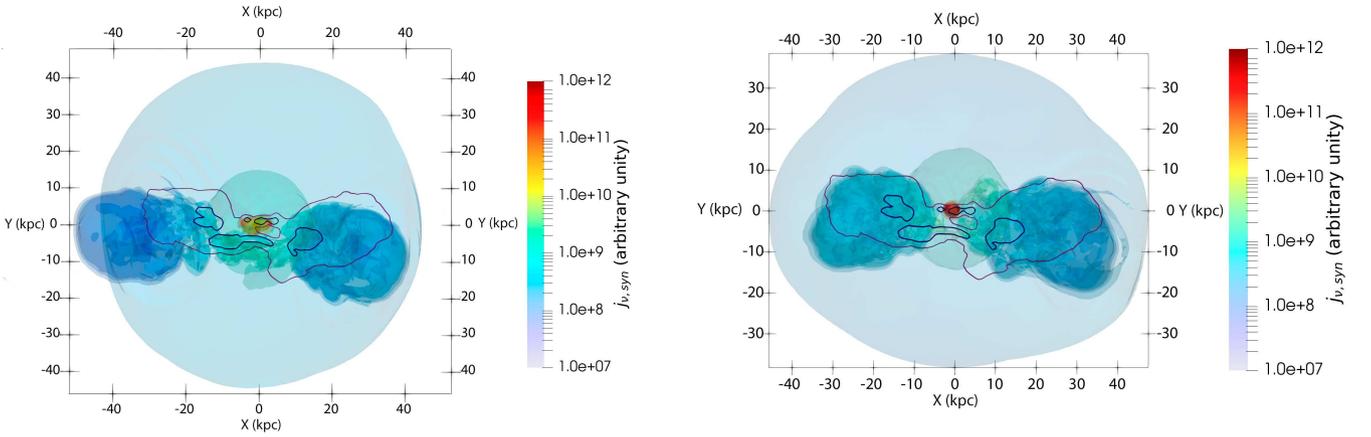

**Figure 14.** Three-dimensional iso-contours (in logarithm scale) of the number density of the last output of the simulation MdEcLt60n8.0v0.1rj1.0. The coloured scale represents the synchrotron emissivity (in arbitrary units) using $\alpha = 1.4$ in Equation 20. Purple and blue contours refer, respectively, to 20 and 200 $\mu$Jy beam$^{-1}$ level intensities of 3C 338 at 4.9 GHz shown in Figure 1.

**Figure 15.** Three-dimensional iso-contours (in logarithm scale) for the simulation QEcLt40n13.0v0.1rj1.0 assuming a time constant jet power and considering the orbital model Q. The coloured scale represents the synchrotron emissivity (in arbitrary units) using $\alpha = 1.4$ in Equation 20. Purple and blue contours refer, respectively, to 20 and 200 $\mu$Jy beam$^{-1}$ level intensities of 3C 338 at 4.9 GHz shown in Figure 1.

We show in Figure 15 the results regarding the Q model through 3D renderings of the pseudo synchrotron emissivity. This model is capable of reproducing the main structures of 3C 338, generating a structure where we can identify the ridge with a curvature close to that observed at 4.9 GHz. However, the restarting jet did not develop as expected (extension smaller than that implied by 200 $\mu$Jy yellow contour in Figure 15. Increasing the power of the jet would cause the structure as a whole to propagate much further than expected. This result is an indication that a scenario with fixed jet power might be insufficient to reproduce the details of the 4.9-GHz map of 3C 338, motivating us to explore simulations with time variable jet power.

**Table 6.** General parameters for the HD simulations with time-dependent jet power performed in this work.

| Simulation ID | $\Delta t_{\rm simul}$ (Myr) | $t_1$ (Myr) | $\tau_1$ (Myr) | $n_1$ ($10^{-4}$ cm$^{-3}$) | $t_2$ (Myr) | $\tau_2$ (Myr) | $n_2/n_1$ | $v''_{\rm jet}$ (c) | $r_{\rm jet}$ (kpc) | $L_{\rm jet}(0)$ ($10^{43}$ erg s$^{-1}$) | $L_{\rm jet}(\Delta t_{\rm simul})$ ($10^{43}$ erg s$^{-1}$) |
|---|---|---|---|---|---|---|---|---|---|---|---|
| MdEvLt60n10.0v0.1rj1.0 | 60 | 0.0 | 20.86 | 10.0 | 60.0 | 1.50 | 1.12 | 0.1 | 1.0 | 67.22 | 75.29 |
| QEvLt40n20.0v0.1rj1.0 | 40 | 0.0 | 11.74 | 20.0 | 42.0 | 4.00 | 0.90 | 0.1 | 1.0 | 11.30 | 0.03 |
| REvLt35n20.0v0.1rj1.0 | 35 | 10.0 | 7.88 | 20.0 | 35.0 | 2.65 | 1.50 | 0.1 | 1.0 | 11.30 | 0.07 |

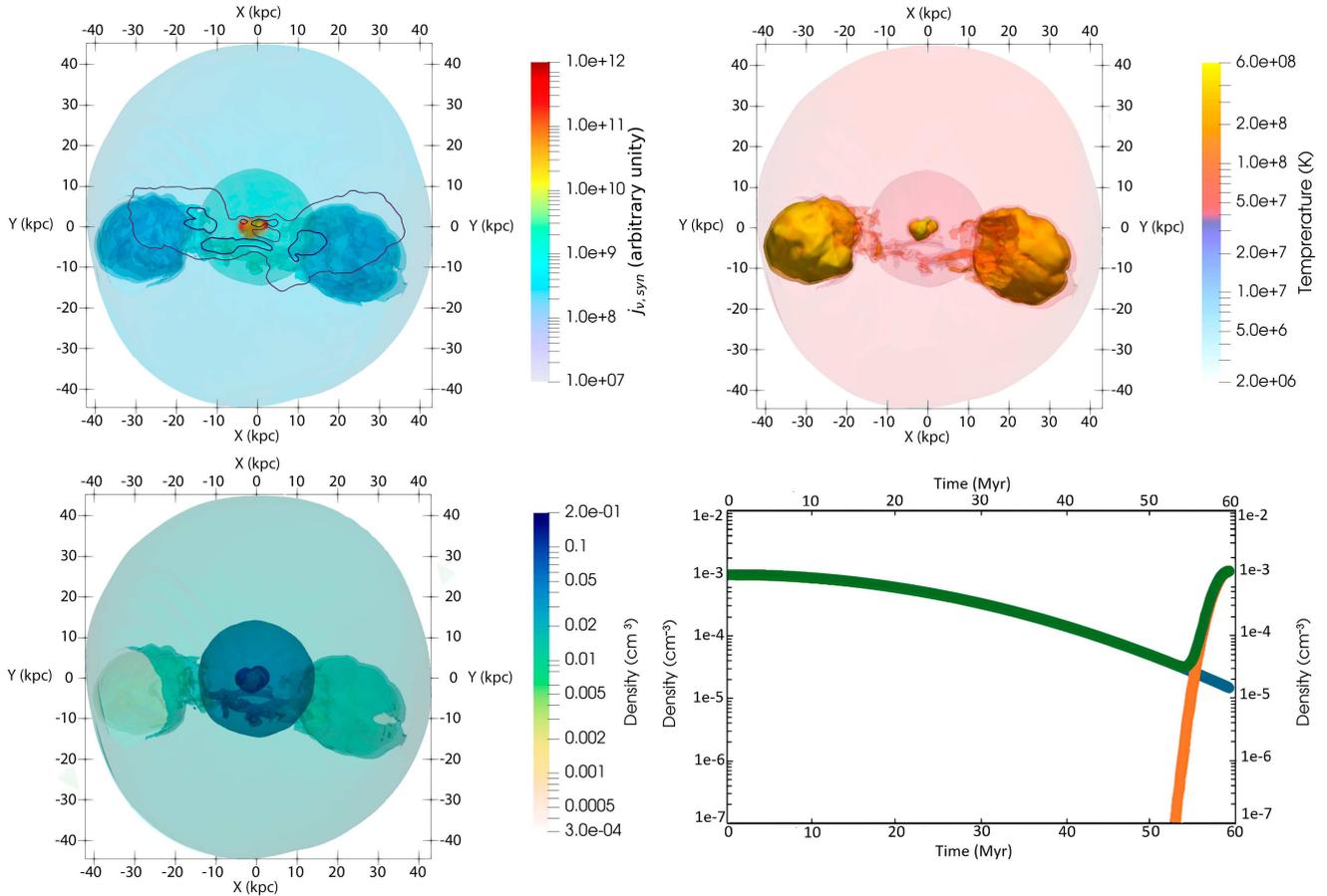

**Figure 16.** Three-dimensional iso-contours (in logarithmic scale) and density variation for the simulation MdEvLt60n10.0v0.1rj1.0, assuming a time-dependent jet power and considering the orbital model Md. In the top left panel, the color scale represents the synchrotron emissivity (in arbitrary units) using $\alpha = 1.4$ in Equation 20. Purple and blue contours correspond to 20 and 200 $\mu$Jy beam$^{-1}$ intensity levels of 3C 338 at 4.9 GHz, shown in Figure 1. In the top right panel, the colour scale represents the temperature. In the bottom left panel, the colour scale represents the number density. The bottom right panel shows the instantaneous jet number density (green line) calculated from Equation 23, which involves the sum of two Gaussian profiles (blue and orange lines).

### 4.2 Time variable jet power simulations

Motivated by observational evidence that the jet power of 3C 338 has changed during the last tens of Myr (e,g,, Allen et al. 2006; Gentile et al. 2007; Nulsen et al. 2013), we decided to relax our previous constant jet power assumption, considering a jet with time-dependent power. As such variations could be generated by changes in the accretion rate onto its supermassive black hole (and leading to fluctuations in the amount of matter channelled to the jet), we have parameterised ad hoc the instantaneous jet number density in terms of sum of two Gaussian functions:

$$n_{\rm jet}(t) = n_1 \left[ e^{-\frac{1}{2}\left(\frac{t-t_1}{\tau_1}\right)^2} + \frac{n_2}{n_1} e^{-\frac{1}{2}\left(\frac{t-t_2}{\tau_2}\right)^2} \right], \quad (23)$$

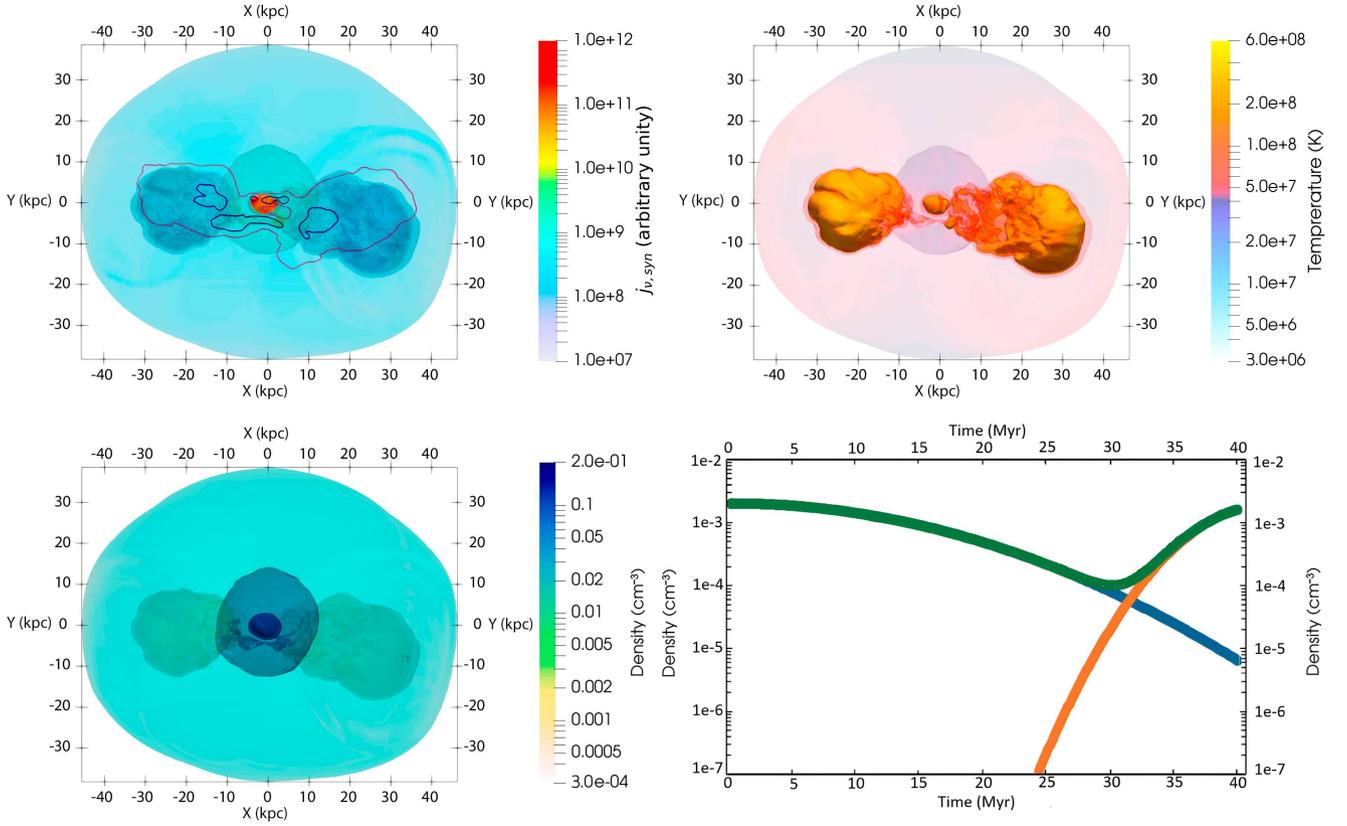

**Figure 17.** Three-dimensional iso-contours (in logarithmic scale) and density variation for the simulation QEvLt40n20.0v0.1rj1.0f1, assuming a time-dependent jet power and considering the orbital model Q. In the top left panel, the color scale represents the synchrotron emissivity (in arbitrary units) using $\alpha = 1.4$ in Equation 20. Purple and blue contours correspond to 20 and 200 $\mu$Jy beam$^{-1}$ intensity levels of 3C 338 at 4.9 GHz, shown in Figure 1. In the top right panel, the colour scale represents the temperature. In the bottom left panel, the colour scale represents the density. The bottom right panel shows the instantaneous jet number density (green line) calculated from Equation 23, which involves the sum of two Gaussian profiles (blue and orange lines).

where $n_1$ and $n_2$ are respectively number densities associated with the first and second Gaussian peaks with widths $\tau_1$ and $\tau_2$ that occur at the epochs $t_1$ and $t_2$.

We show in Table 6 the numerical setup for the time variable jet power simulations performed in this work. Two of these models are based on the previous simulations MdEcLt60n8.0v0.1rj1.0 and QEcLt40n13.0v0.1rj1.0 with a time-constant jet power. The instantaneous jet number density calculated from Equation 23 and implemented in our non-constant jet power simulations are shown in Figure 16, Figure 17 and Figure 18. The last output from simulation MdEvLt60n10.0v0.1rj1.0 is presented in Figure 16. The extension of outer parts of the simulated jet agrees with the size of the East and West lobes at 4.9 GHz (Figure 1). A ridge-like structure is also generated in this simulation, agreeing in location and size with the radio ridge detected in 3C 338. However, the observed ridge's curvature is not fully reproduced in this simulation, probably due to the combination between a relatively low galactic velocity ($\sim$ 150 km s$^{-1}$) and a moderate jet number density ($\sim 10^{-4}$ cm$^{-3}$) at the ridge's position. Concerning the restarting jet, it did not propagate the whole projected distance seen in the radio maps of 3C 338, even for the relatively high jet number density (between $\sim 2 \times 10^{-4}$ and $1.1 \times 10^{-3}$ cm$^{-3}$) employed during the last 3 Myr of evolution.

The simulation MdEvLt60n10.0v0.1rj1.0 was evolved for 60 Myr, an interval longer than the synchrotron ages of the outer lobes of 3C 338 inferred by Burns et al. (1983) ($\sim$ 30 Myr). We show in Figure 17 the 3D rendering of the pseudo synchrotron emissivity, temperature and density maps of the simulation QEvLt40n20.0v0.1rj1.0 after evolving the jet for 40 Myr. The outermost structures have dimensions consistent with those of the East and West lobes at 4.9 GHz, also producing a ridge structure with curvature similar to that observed in the 4.9-GHz interferometric images of 3C 338. However, the restarting counter-jet did not develop sufficiently, propagating about a half of the projected distance between the core and the western inner lobe at 4.9 GHz (no relevant discrepancy is noted between the size of the simulated jet and the eastern inner lobe).

Aiming to improve our previous results, we tilted clockwise the orbit of the model Q and reduced its $V_{\rm tot}'''$ to $\sim$ 555 km s$^{-1}$, which led the new orbit to cross the brightest region of the ridge. This new orbital model labelled R (see Table 1 and Table 3 for its main characteristics) was used to carry out the simulation REvLt35n20.0v0.1rj1.0. Its jet parameters are listed in Table 6. The last output from the simulation REvLt35n20.0v0.1rj1.0 after 35 Myr of evolution is shown in Figure 18. Although the ridge structure has become more diffuse in comparison with simulation QEvLt40n20.0v0.1rj1.0, the restarting jet evolved as expected, reaching the projected sizes of the eastern and western inner lobes at 4.9 GHz. Simulation REvLt35n20.0v0.1rj1.0 is capable of reproducing the overall shape of the extended structures and ridge, as well as their correct locations, but a slight inclination of the outer lobe-like structures to the south still remains, even though much less evident than that observed in simulation QEvLt40n20.0v0.1rj1.0.

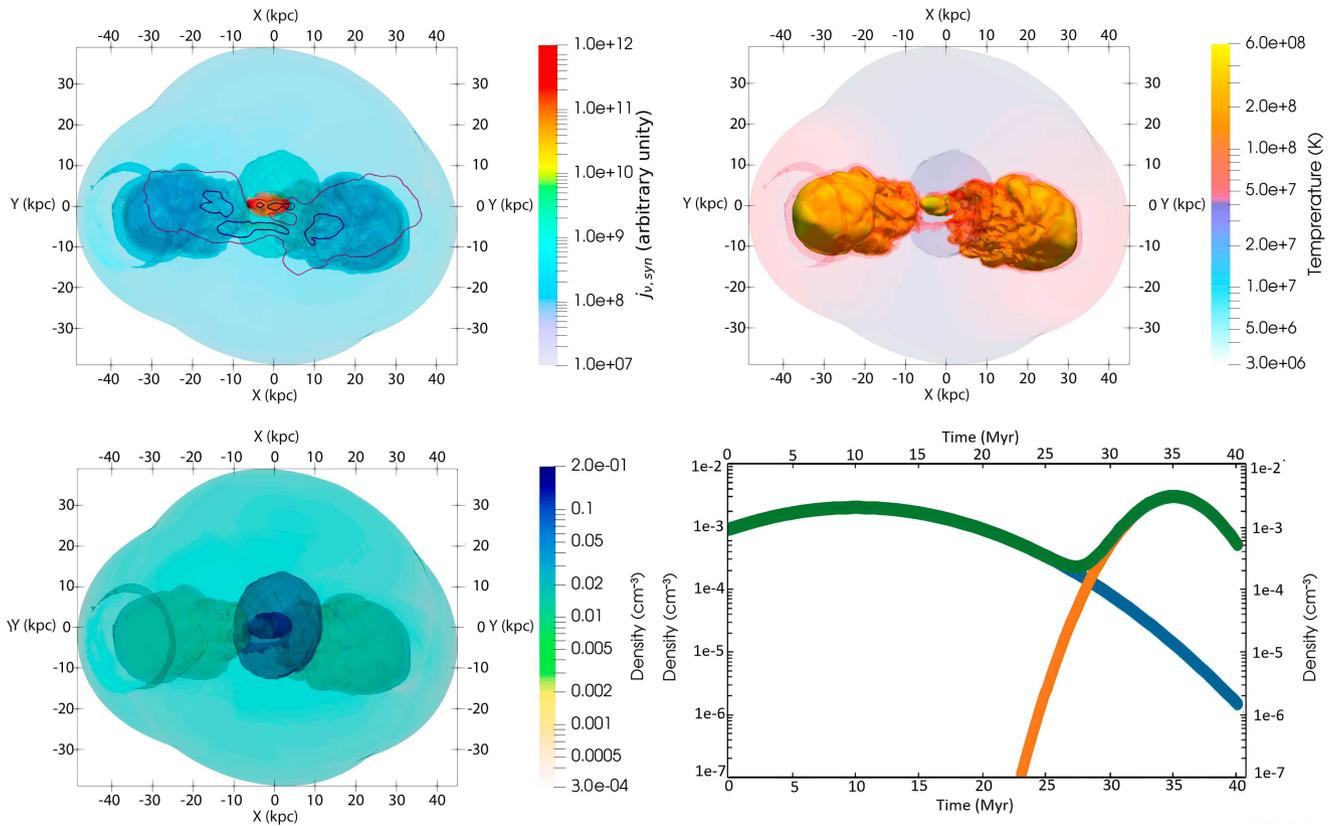

**Figure 18.** Three-dimensional iso-contours (in logarithmic scale) and density variation for the simulation REvLt35n20.0v0.1rj1.0, assuming a time-dependent jet power and considering the orbital model R. In the top left panel, the colour scale represents the synchrotron emissivity (in arbitrary units) using $\alpha = 1.4$ in Equation 20. Purple and blue contours correspond to 20 and 200 $\mu$Jy beam$^{-1}$ intensity levels of 3C 338 at 4.9 GHz, shown in Figure 1. In the top right panel, the colour scale represents the temperature. In the bottom left panel, the colour scale represents the density. The bottom right panel shows the instantaneous jet number density (green line) calculated from Equation 23, which involves the sum of two Gaussian profiles (blue and orange lines).

Thus, simulations with a time-dependent jet power (specially simulation REvLt35n20.0v0.1rj1.0) seems to be favoured among all the simulations performed in this work. Indeed, observational evidence that the jet power of 3C 338 has varied over the years (e.g., Allen et al. 2006; Gentile et al. 2007; Nulsen et al. 2013) supports our initiative in exploring numerically the impact of a time-changing jet power on the radio morphology of this source. The small issues mentioned above may be a consequence of the lack of any ICM anisotropy in our simulations, since our models consider an initial symmetrically radial distribution for DM and ICM gas (see section 5).

## 5 DISCUSSION

### 5.1 The ridge feature in 3C 338

As the power emitted by a relativistic electron moving through a magnetised region is proportional to the square of its own total energy (e.g., Pacholczyk 1970), high energy electrons loose their kinetic energy at faster rates, steepening their initial (power-law) energy distribution at higher frequencies. The consequence of this process is producing a steeper continuum spectrum at higher radio frequencies. Thus, spectral index measurements along extended radio sources can be used as a probe for detecting age gradients across it. In the case of 3C 338, its radio emission above $\sim$ 1 GHz is compatible with an optically-thin synchrotron spectrum, besides its kiloparsec-scale jet presenting substantial changes in $\alpha$ along the source (e.g., Burns et al. 1983). Values of $\alpha$ between 1.47 and 4.87 GHz were derived for the main structures of 3C 338 (see Figure 1), with the steepest values found in its outskirts (East and West lobes), mid values in the southern structures (ridge and lobes 1 and 2), and flatter ones in the new jet region (nowadays core and Eastern and Western inner jets).

Burns et al. (1983) estimated a synchrotron lifetime of about 8 Myr for the ridge, 9 Myr for the lobe 1, and 30 Myr for the East lobe. In addition, Nulsen et al. (2013) estimated a syncrotron age of $\sim$ 5 Myr for the lobe-like structure in the end of the restarting jets, reinforcing their younger nature compared to the ridge and the other main structures detected in 3C 338. Note that ageing calculations based on spectral steepening due to synchrotron or inverse Compton processes may provide inaccurate estimates of the age's source due to several potential issues, such as the presence of magnetic-field gradients along the source and strong jet deceleration (e.g., Blundell & Rawlings 2000, 2001; Rudnick 2002). In this case, longer dynamical timescales are necessary for the jets reaching their observed projected sizes, which translates to a lower limit for the source's age when the spectral ageing method is used (e.g., Rudnick 2002). It motivated us to adopt longer ages for the East and West lobes in some HD simulations performed in this work compared to those derived by Burns et al. (1983). Notwithstanding, the spectral ageing estimates for the lifetime of the ridge and the current jet (Eastern and Western inner lobes in Figure 1) by Burns et al. (1983) fairly agree with the estimates by Nulsen et al. (2013), who used a different approach in their calculations. This agreement suggests that potential issues regarding spectral ageing mentioned previously might be negligible at least for the ridge and the restarting jet in 3C 338.

The presence of age gradients across 3C 338, together with the jet-like shape of the kiloparsec-scale ridge suggest that some odd phenomenon has been happening in the centre of A 2199 during the last tens of Myr. Burns et al. (1983) invoked two different scenarios for explaining the unusual ridge structure: ram-pressure dragging forces due to a cooling accretion flow onto NGC 6166 that swept out an older jet produced at the actual core's position to the current ridge's location, and the motion of the radio core either around the baricentre of NGC 6166 and NGC 6166B or around the gravitational potential well of A 2199.

Burns et al. (1983) found that the ram pressure due to a cooling spherical accretion flow is on the same order of the minimum value of the internal pressure of the ridge, necessary condition to avoid its collapse during its journey to its actual position. However, these authors pointed out some potential issues regarding this scenario: this past accretion flow must be asymmetric (more intense towards the south) to drag former jet plasma to the current ridge's position; its steep radio spectrum ($\alpha \sim 1.7$) is not characteristic of a shocked region; no limb brightening is seen on the southern edge of the ridge (the supposed contact surface between the dragged ridge and the ICM). Nulsen et al. (2013) also argue in favour of a ram-pressure origin for the radio ridge, but proposing that it is an older jet plasma removed from the core by a sloshing cold front induced by the passage of a merging subcluster about 400 Myr ago. Indeed, X-ray observations by Nulsen et al. (2013) and recent SPH simulations by Machado et al. (2022) strongly supports the existence of a sloshing phenomenon in A 2199. To preserve the jet morphology of the ridge during ~ 8 Myr, a non turbulent gas flow with a speed of about 800 km s$^{-1}$ is necessary (Nulsen et al. 2013). However, such value is too high when compared with a typical speed of a sloshing front (e.g., Ascasibar & Markevitch 2006). It is important to emphasise that the possibility of the ridge has been induced by a sloshing front is out of the scope of the analyses conducted in this work. Dedicated HD simulations including the sloshing cold fronts should be pursued to check the viability of this scenario in 3C 338 in future works.

The second possibility about the physical origin of the ridge raised by Burns et al. (1983) is based on the motion of the radio core of NGC 6166 within the cD halo, so that the ridge would be an older jet produced in a displaced location about 8 Myr ago. Observational evidence supporting this scenario comes from the similarity between jets in radio galaxies and the shape of the ridge, a well collimated and symmetrical (along its transverse direction) structure with its extremities ending in the radio lobes 1 and 2 (Burns et al. 1983). We would like to include an additional observational feature to this supporting list: the projected curvature of the ridge. As highlighted by the round white arc in Figure 1, the radio morphology of the system ridge–lobe 1–lobe 2 resembles that seen in the Wide Angle Tail (WAT) radio galaxies (e.g., Owen & Rudnick 1976; O'Dea & Baum 2023). WAT sources are typically found in more dominant cluster galaxies, showing jets that are bent due to their interaction with the ICM (e.g., Owen & Rudnick 1976; Burns 1986; O'Donoghue et al. 1993; Bird 1994; Sakelliou & Merrifield 2000; Missaglia et al. 2023). Indeed, NGC 6166 is the BCG galaxy of A 2199, which corroborates such interpretation.

Furthermore, Vallee et al. (1981) showed that the unusual radio morphology of the head-tail radio galaxy IC 708 could be interpreted as projection effects due to changes in its orbital velocity on the jet plasma. In contrast with what is expected from galactic displacements occurring always at same direction, an inversion in the sense of the motion can reverse the orientation of the jet's curvature, as shown in figure 5b in Vallee et al. (1981). Such inversions in the sense of propagation can be produced by a curved trajectory oriented close to the line of sight. The unusual jet morphology of the radio galaxy IC 708 is surprisingly similar to the system ridge–lobes 1–lobe 2 in 3C 338. Except for model Ma, all orbital motions considered in this work have an inversion in the sense of the motion (roughly from North to South to South to North) occurring in a region close to the ridge, necessary condition to bend a jet (ridge) that ends to two approximate hook-shaped radio structures, like lobes 1 and 2 in Figure 1.

### 5.2 The motion of the jet inlet region

In this work, we performed the first 3D HD simulations in the literature aiming to verify whether a non-static radio core could reproduce the complex radio morphology of 3C 338 at kiloparsec scales, including the peculiar radio ridge. In contrast to Burns et al. (1983), who considered analytical models using either a circular orbit about the barycentre of the cD nuclei or an oscillation about the centre of the cluster gravitational potential well, we have assumed closed orbits of the radio core around the X-ray brightness peak of A 2199, the supposed centre of this galaxy cluster. The right ascension and declination offsets between the nucleus of NGC 6166 and the X-ray brightness peak of A 2199, the radial velocity of NGC 6166 and the inferred DM profile for A 2199 were used as observational constraints to the integration of the radio core's orbit around the cluster centre (see section 2 for details). The orbits were numerically integrated back in time using different intervals: from 35 to 60 Myr, with the longest interval corresponding to twice the estimated age of the oldest structures seen in Figure 1.

Regarding the current velocity of NGC 6166 predicted by the orbital scenarios tested in this work, only $V''_{\rm tot} \gtrsim 450$ km s$^{-1}$ are allowed since lower values lead to jet structures that have ages incompatible with the synchrotron lifetimes estimated by Burns et al. 1983. Note that this lower limit of about 450 km s$^{-1}$ could be relaxed if the spectral age of the East and West lobes were underestimated due to the issues concerning ageing calculations (see previous discussion in subsection 5.1). The maximum value for $V''_{\rm tot}$ considered in this work was ~ 798 km s$^{-1}$, about 97% of the A 2199's cluster velocity dispersion inferred by Bender et al. (2015). Such (3D) velocities are relatively high when compared with typical (1D) velocities of cD galaxies in relaxed clusters, but not so different in relation to BCG galaxies in clusters exhibiting signatures of merger and/or collision with other groups or clusters, where galaxies present peculiar radial velocities as fast as 1000 km s$^{-1}$ (e.g., Coziol et al. 2009; Lopes et al. 2018; De Propris et al. 2021). Indeed, A 2199 is part of a supercluster where neighbouring cluster A 2197 and other groups of galaxies are probably falling into it (e.g., Rines et al. 2001, 2002), showing also a sloshing feature seen in the residual X-ray images of its central region (Nulsen et al. 2013). These characteristics indicate that A 2199 is a non-relaxed cluster where past dynamical interactions could have driven galaxy motions similar to those implemented in our HD simulations.

Machado et al. (2022) showed numerically that a non-frontal collision with a galaxy group of $M_{200} = 1.6 \times 10^{13}$ M$_\odot$ can shift the density/gravitational potential DM peak of A 2199 in relation to their initial (and coincident) location, introducing displacements reaching about 40 kpc. Such effect, first discussed in Ascasibar & Markevitch (2006), has probably impacted NGC 6166 too, dragging it together with the DM peak. In the case of the standard non-relativistic (cold), collisionless dark matter with a pure cuspy (NFW) profile, none or small-amplitude oscillations ($\lesssim 2$ kpc; Harvey et al. 2017) of the BCG galaxy around the bottom of the DM gravitational potential well are expected since it is tightly bound at the DM peak (Schaller et al. 2015; see also figure D1 in Kim et al. 2017). However, a cored density DM profile (similar to Equation 3) or even alternative forms of DM (e.g., self-interacting dark matter Yoshida et al. 2000) allow BCG galaxies to wobble when perturbed by a merger involving another cluster or group of galaxies (e.g., Kim et al. 2017; Harvey et al. 2017, 2019). Therefore, given the apocentric radius of our most promising orbital models are smaller than about 7 kpc, there is some room for a possible wobbling motion of NGC 6166 even considering a standard cold DM profile for A 2199.

The HD simulations presented in this work involve the motion of the jet inlet region following orbits previously integrated under observational constraints for NGC 6166 (3C 338) and A 2199 (see section 2). It means we have assumed implicitly that NGC 6166 is orbiting the centre of A 2199. However, this is not the only option. As pointed out by Chu et al. (2023), the central supermassive black hole (SMBH) of a BCG can suffer gravitational kicks by dynamical interactions with satellite galaxies, gaining enough kinetic energy to put it at distances as large as hundreds of kiloparsecs, which is quite farther than the largest apocentric radius used in this work (∼ 14 kpc; see Table 3). Thus, it might be possible that the same group-cluster encounter that originated the sloshing pattern seen at X-ray wavelengths has also provided a gravitational kick to the accreting SMBH. The wandering SMBH scenario could also better accommodate the jet power time variation in 3C 338 (e,g,, Allen et al. 2006; Gentile et al. 2007; Nulsen et al. 2013) since the accretion rate onto the SMBH would vary accordingly to the instantaneous position of the SMBH (higher accretion rates occurring in the central region of galaxies where the denser gas reservoirs are usually located; Chu et al. 2023). Interestingly, there is a steep brightness decrement between the locations of the ridge and the new jet in Figure 1, which might be interpreted as an indirect evidence of a past decrease of the jet activity between the formation of the ridge structure and the current jet). However, it is worthy to note that a straightforward link between radio brightness and intrinsic jet power is not an easy task, since dissipative processes along the jet evolution may mimic a past decrease of the later quantity.

### 5.3 The sloshing front in A2199

Finally, it was not considered in this work non-radial anisotropies in the initial spatial distributions of the number density and temperature of the ICM due to a sloshing front. The main reason for that was to avoid introducing extra parameters (not known a prior) into our simulations, keeping them as simple as possible. However, as the sloshing phenomenon creates contact discontinuities in the gas density and temperature (but not in the thermal pressure), it might have some impact on the jet propagation.

We show in Figure 19 the X-ray residual map of A 2199 obtaibed by Machado et al. (2022) after subtracting a 2D $\beta$-model from the XMM-Newton X-ray data. White contours that delineates 3C 338 at 4.9 GHz are also superposed to the residual map, which reveals a spiral-like structure (sloshing front) in the central region of A 2199. The outer parts of 3C 338 end roughly at the external "walls" of the sloshing front, suggesting some role of the contact discontinuities in confining the outer lobes of this radio galaxy. Besides, there is a clear (asymmetric) enhancement of the residuals towards the actual radio core of 3C 338 (about a factor of 2 between the ridge and the core locations), suggesting an increase of the gas density towards the active radio core of NGC 6166. It is supported by optical observations of NGC 6166, which reveal several extended dust filaments close to its nucleus (Martel et al. 2004; see also figure A6 in Tremblay et al. 2015), besides the A 2199 is a cool-core galaxy cluster with a strong cooling flow operating in its central region (∼ 170 M$_\odot$ yr$^{-1}$; Peres et al. 1998).

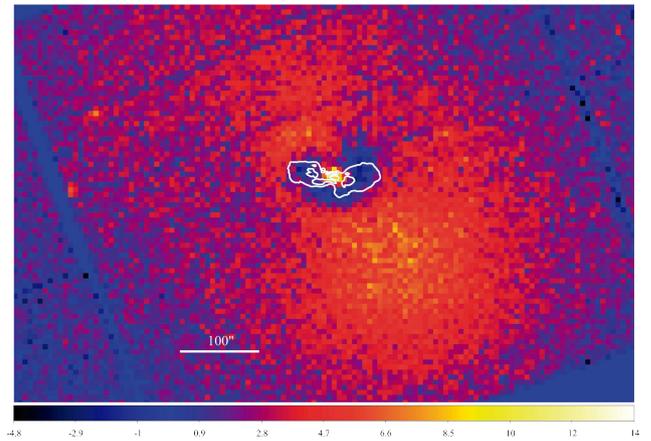

**Figure 19.** Residual image obtained by subtracting the 2D $\beta$ model from the surface brightness image of Abell 2199. Obtained after analysis of XMM-Newton X-ray data by Machado et al. (2022). White contours have the same meaning of those in Figure 7.

The increase of the gas density towards the AGN at kiloparsec scales seen in Figure 19 does not necessarily imply an instantaneous increment of the accretion rate onto the SMBH (at sub-parsec scales). Indeed, how the gas looses its angular momentum at galactic scales to be accreted onto a SMBH is still a matter of debate in the literature (e.g, Combes 2021). However, the mentioned increase of gas density argues in favour of a non-static gas reservoir, which will probably translates to a subsequent time-dependent SMBH accretion rate in NGC 6166. Therefore, the current jet power of 3C 338 must be related to older episodes of gas infall faced by NGC 6166 during its evolution, where its current weaker jet power could reflect a low-mass infall of gas occurred million of years ago.

## 6 CONCLUSIONS

In this work, we performed the first 3D HD simulations in the literature aiming to verify whether a non-static radio core could reproduce the complex radio morphology of 3C 338 at kiloparsec scales, including a peculiar feature labelled as ridge. In contrast to Burns et al. (1983), who considered analytical models using either a circular orbit about the barycentre of the cD nuclei or an oscillation of NGC 6166 about the centre of the cluster gravitational potential well, we have assumed closed orbits of the radio core around the X-ray brightness peak of A 2199, the supposed centre of this galaxy cluster.

The physical characteristics of the intergalactic medium in which simulated jet and counter-jet propagate were constrained from the X-ray observations in the vicinity of NGC 6166. Possible orbits for the jet inlet region were derived from the measured radial peculiar velocity of NGC 6166, as well as the angular offset between its position and the X-ray peak on the plane of the sky, while the jet parameters were constrained by parsec-scale interferometric radio observations (Feretti et al. 1993; Gentile et al. 2007; Yuan et al. 2018) and the estimated jet power of 3C 338 obtained from radio and X-ray data set (e.g., Allen et al. 2006; Gentile et al. 2007; Nulsen et al. 2013).

All the HD simulations conducted in this work involved a pair of jets injected from a cylindrical jet inlet region that follows an orbit previously integrated under a DM gravitational potential composed of two components (e.g., Hogan et al. 2017): a more extended NFW profile and a more compact cored, isothermal profile. Those jets were evolved during tens of million years (35-60 Myr, depending on the simulation) to respect the synchrotron lifetimes estimated by the different radio structures in 3C 338.

The main results achieved in this work are summarised as follows:

- Orbits with current velocities lower than $\sim 450$ km s$^{-1}$ are ruled out since the simulated East and West lobes would have ages incompatible (longer than a factor of two) with their synchrotron lifetimes estimated by Burns et al. (1983). Note that this limit could be lowered if the ages of the East and West lobes were underestimated due to the issues concerning the spectral ageing method (e.g., Blundell & Rawlings 2000, 2001; Rudnick 2002).;

- An orbital plane seen at angles $\theta$ lesser than 14°(faced-on orbits) are strongly disfavoured because the current velocity of NGC 6166 would be higher than about 819 km s$^{-1}$ (the A 2199's cluster velocity dispersion; Bender et al. 2015). However, our simulations were not able to put strict constraints upon the value of $\theta$ when the time constant jet power condition is relaxed. Regarding the rotation angle $\varphi$, a narrow range of values is allowed for each orbital model considered in this work ($\Delta\varphi \lesssim 30°$);

- To respect the age gradient in 3C 338, as well as the apocentric passage taking place near the ridge's location, the orbital motion must occur in a clockwise sense around the X-ray peak in A 2199;

- Except for the model Ma, all orbital motions considered in this work have an inversion in the sense of the motion (roughly from North to South to South to North) occurring in a region close to the ridge. Indeed, this is a necessary condition to bend a jet (ridge) that ends to two approximate hook-shaped radio structures , like lobes 1 and 2 in Figure 1 (similar to the head-tail radio galaxy IC 708; Vallee et al. 1981). It reinforces the interpretation that the projected curvature of the ridge in 3C 338 at 4.9 GHz would be due to interaction with the ICM as the jet moves through it, as in the case of WAT radio galaxies (e.g., Owen & Rudnick 1976; O'Dea & Baum 2023);

- All simulations in their last output have their brightest region coinciding with the current position of the active core of NGC 6166 and its younger jet/counter-jet seen in the 4.9-GHz map of 3C 338;

- Simulations with a jet inlet region with a radius of 2 kpc were used to explore the overall dependence of the jet and/or orbital parameters on the jet evolution. Under the assumption of time constant jet power, the simulated outer lobes tend to curve towards North direction, with the highest bends being observed in the simulations with higher orbital speeds (e.g., model J), a clear effect of ram-pressure forces due to the relative motion between the AGN and the ICM (e.g, Gunn & Gott 1972; Begelman et al. 1979; Vallee et al. 1981; Baan & McKee 1985). The E-W projected lengths of the simulated jets also exhibit a general trend where a jet pointing West are slightly longer than those with a jet oriented easterly, whatever the orbital model adopted in the simulation. The reason for that is the differences between the values of $\epsilon_{jet}$ (angle between the jet and the $z''$-axis, which is perpendicular to the orbital plane) obtained in the East and West jet scenarios: for a given orbital model, $\epsilon_{jet}$ is always larger for East-jet solutions, implying that the jet is closer to the orbital plane and more susceptible to strong interactions with the surrounding ambient due to the galactic motion around the A 2199's centre (e.g., Baan & McKee 1985 and Appendix C). The same reason is behind of the counter intuitive result where the increase of $\vartheta_{jet}$ made the projected jet structures more compact;

- The decrement of the radius of the jet inlet region by a factor of two (from 2 kpc to 1 kpc) did not influenced the overall shape of the simulated jets, suggesting the achievement of numerical convergence in our results;

- Although the outer parts of 3C 338 can be easily reproduced in our HD simulations, the simultaneous recovering of the ridge structure (in intensity and curvature) and the projected extension of the eastern and western inner lobes is not trivial. An illustrative example is the simulation QEcLt40n13.0v0.1rj1.0, in which a ridge feature with the a curvature close to that observed at 4.9 GHz is formed, but the restarting jet did not propagated as expected, suggesting that scenarios adopting time varying jet power must be considered in the case of 3C 338. Indeed, there is observational evidence that its jet power has changed during the last tens of Myr (e,g,, Allen et al. 2006; Gentile et al. 2007; Nulsen et al. 2013). Note that our simulation REvLt35n20.0v0.1rj1.0 (see Figure 18) with a time-dependent jet power fairly recovered the main jet structures present in Figure 1 at expected timescales.

Regarding the current (3D) velocities of NGC 6166 predicted by the orbital scenarios tested in this work are relatively high when compared with typical (1D) velocities of BCG galaxies in relaxed clusters, but not so different in relation to those in clusters exhibiting signatures of merger and/or collision with other groups or clusters (e.g., Coziol et al. 2009; Lopes et al. 2018; De Propris et al. 2021). Indeed, A 2199 is part of a supercluster where neighbouring cluster A 2197 and other groups of galaxies are probably falling into it (e.g., Rines et al. 2001, 2002), showing also a sloshing feature seen in the residual X-ray images of its central region (Nulsen et al. 2013). These characteristics indicate that A 2199 is a non-relaxed cluster where past dynamical interactions could have driven galaxy motions similar to those implemented in our HD simulations.

Machado et al. (2022) showed numerically that a non-frontal collision with a galaxy group of $M_{200} = 1.6 \times 10^{13}$ M$_\odot$ can shift the density/gravitational potential DM peak of A 2199 in relation to their initial (and coincident) location, introducing displacements reaching about 40 kpc. First discussed in Ascasibar & Markevitch (2006), such effect has probably impacted NGC 6166 too, dragging it together with the DM peak (see section 5 for further details).

It is important to emphasise that the dynamical scenario where NGC 6166 moves around the X-ray inferred centre of A 2199 (or even a wandering SMBH that suffered gravitational kicks by dynamical interactions with satellite galaxies; Chu et al. 2023) is not only the unique possibility for explaining the ridge structure in 3C 338. Ram-pressure dragging forces due to either a cooling accretion flow onto NGC 6166 (Burns et al. 1983) or a sloshing cold front induced by the passage of a merging subcluster 400 Myr ago (Nulsen et al. 2013) could have swept out an older jet produced at the actual core's position to the current ridge's location. Indeed, X-ray observations by Nulsen et al. (2013) and recent SPH simulations by Machado et al. (2022) strongly supports the existence of a sloshing phenomenon in A 2199.

Specific HD simulations including sloshing cold fronts should be pursued to check the viability of this scenario in 3C 338 in future works. Moreover, the inclusion of magnetic fields into future simulations is desirable to verify their role in shaping the radio jet of 3C 338, as well as providing synthetic polarisation maps that can be compared to those found in the literature (e.g., Vacca et al. 2012).

**ACKNOWLEDGEMENTS**

The authors acknowledge the National Laboratory for Scientific Computing (LNCC/MCTI, Brazil) for providing HPC resources of 20   *A. S. R. Antas et al.*the SDumont supercomputer, which have contributed to the research results reported within this paper. URL: http://sdumont.lncc.br. ASRA, REGM and TFL acknowledge the financial support from Conselho Nacional de Desenvolvimento Científico e Tecnológico CNPq through grants 684846/2022-00, 307205/2021-5 and 303672/2022-6 respectively. REGM acknowledges support from *Fundação de Apoio à Ciência, Tecnologia e Inovação do Paraná* through grant 18.148.096-3 – NAPI *Fenômenos Extremos do Universo*. We would like to thank the anonymous referee for the constructive comments and insightful suggestions.**DATA AVAILABILITY**

The data generated in this research will be shared on reasonable request to the corresponding author.

**REFERENCES**

Allen S. W., Dunn R. J. H., Fabian A. C., Taylor G. B., Reynolds C. S., 2006, MNRAS, 372, 21
Ascasibar Y., Markevitch M., 2006, ApJ, 650, 102
Baan W. A., McKee M. R., 1985, A&A, 143, 136
Begelman M. C., Rees M. J., Blandford R. D., 1979, Nature, 279, 770
Bender R., Kormendy J., Cornell M. E., Fisher D. B., 2015, ApJ, 807, 56
Binney J., Tremaine S., 1987, Galactic dynamics. Princeton University Press
Bird C. M., 1994, AJ, 107, 1637
Blundell K. M., Rawlings S., 2000, AJ, 119, 1111
Blundell K. M., Rawlings S., 2001, in Laing R. A., Blundell K. M., eds, Astronomical Society of the Pacific Conference Series Vol. 250, Particles and Fields in Radio Galaxies Conference. p. 363
Bovy J., 2015, ApJS, 216, 29
Burbidge E. M., 1962, ApJ, 136, 1134
Burns J. O., 1986, Canadian Journal of Physics, 64, 373
Burns J. O., Schwendeman E., White R. A., 1983, ApJ, 271, 575
Caproni A., Lanfranchi G. A., Friaça A. C. S., Soares J. F., 2023, ApJ, 944, 11
Chu A., Boldrini P., Silk J., 2023, MNRAS, 522, 948
Combes F., 2021, Active Galactic Nuclei: Fueling and Feedback, doi:10.1088/2514-3433/ac2a27.
Coziol R., Andernach H., Caretta C. A., Alamo-Martínez K. A., Tago E., 2009, AJ, 137, 4795
David L. P., 1997, ApJ, 484, L11
De Propris R., et al., 2021, MNRAS, 500, 310
Edge D. O., Shakeshaft J. R., McAdam W. B., Baldwin J. E., Archer S., 1959, Mem. RAS, 68, 37
Feretti L., Comoretto G., Giovannini G., Venturi T., Wehrle A. E., 1993, ApJ, 408, 446
Ge J., Owen F. N., 1994, AJ, 108, 1523
Gentile G., Rodríguez C., Taylor G. B., Giovannini G., Allen S. W., Lane W. M., Kassim N. E., 2007, ApJ, 659, 225
Giovannini G., Cotton W. D., Feretti L., Lara L., Venturi T., 1998, ApJ, 493, 632
Gordon D., et al., 2016, AJ, 151, 154
Gower A. C., Gregory P. C., Unruh W. G., Hutchings J. B., 1982, ApJ, 262, 478
Gunn J. E., Gott J. Richard I., 1972, ApJ, 176, 1
Harvey D., Courbin F., Kneib J. P., McCarthy I. G., 2017, MNRAS, 472, 1972
Harvey D., Robertson A., Massey R., McCarthy I. G., 2019, MNRAS, 488, 1572
Hogan M. T., McNamara B. R., Pulido F., Nulsen P. E. J., Russell H. R., Vantyghem A. N., Edge A. C., Main R. A., 2017, ApJ, 837, 51
Horton M. A., Krause M. G. H., Hardcastle M. J., 2020, MNRAS, 499, 5765
Hudson D. S., Mittal R., Reiprich T. H., Nulsen P. E. J., Andernach H., Sarazin C. L., 2010, A&A, 513, A37
Johnstone R. M., Allen S. W., Fabian A. C., Sanders J. S., 2002, MNRAS, 336, 299
Kim S. Y., Peter A. H. G., Wittman D., 2017, MNRAS, 469, 1414
Krause M. G. H., et al., 2019, MNRAS, 482, 240
Lachieze-Rey M., Vigroux L., Souviron J., 1985, A&A, 150, 62
Lauer T. R., 1986, ApJ, 311, 34
Lauer T. R., Postman M., Strauss M. A., Graves G. J., Chisari N. E., 2014, ApJ, 797, 82
Li S., 2005, Journal of Computational Physics, 203, 344
Lopes P. A. A., Trevisan M., Laganá T. F., Durret F., Ribeiro A. L. B., Rembold S. B., 2018, MNRAS, 478, 5473
MacDonald G. H., Kenderdine S., Neville A. C., 1968, MNRAS, 138, 259
Machado R. E. G., Laganá T. F., Souza G. S., Caproni A., Antas A. S. R., Mello-Terencio E. A., 2022, MNRAS, 515, 581
Martel A. R., et al., 2004, AJ, 128, 2758
Mignone A., Bodo G., Massaglia S., Matsakos T., Tesileanu O., Zanni C., Ferrari A., 2007, ApJS, 170, 228
Miley G. K., Perola G. C., van der Kruit P. C., van der Laan H., 1972, Nature, 237, 269
Minkowski R., 1958, PASP, 70, 143
Minkowski R., 1961, AJ, 66, 558
Mioduszewski A. J., Hughes P. A., Duncan G. C., 1997, ApJ, 476, 649
Mirakhor M. S., Walker S. A., 2020, Monthly Notices of the Royal Astronomical Society, 497, 3943
Miralda-Escude J., 1991, ApJ, 370, 1
Missaglia V., et al., 2023, A&A, 674, A191
Musoke G., Young A. J., Molnar S. M., Birkinshaw M., 2020, MNRAS, 494, 5207
Nandi S., Caproni A., Kharb P., Sebastian B., Roy R., 2021, ApJ, 908, 178
Navarro J. F., Frenk C. S., White S. D. M., 1997, ApJ, 490, 493
Nawaz M. A., Bicknell G. V., Wagner A. Y., Sutherland R. S., McNamara B. R., 2016, MNRAS, 458, 802
Nulsen P. E. J., et al., 2013, ApJ, 775, 117
O'Dea C. P., Baum S. A., 2023, Galaxies, 11, 67
O'Dea C. P., Owen F. N., 1985, AJ, 90, 954
O'Donoghue A. A., Eilek J. A., Owen F. N., 1993, ApJ, 408, 428
Oegerle W. R., Hill J. M., 2001, AJ, 122, 2858
Owen F. N., Rudnick L., 1976, ApJ, 205, L1
Pacholczyk A. G., 1970, Radio astrophysics. Nonthermal processes in galactic and extragalactic sources. W.H.Freeman & Co Ltd
Parker E. A., Kenderdine S., 1967, The Observatory, 87, 124
Peres C. B., Fabian A. C., Edge A. C., Allen S. W., Johnstone R. M., White D. A., 1998, MNRAS, 298, 416
Perucho M., Martí J.-M., Quilis V., Borja-Lloret M., 2017, MNRAS, 471, L120
Rines K., Mahdavi A., Geller M. J., Diaferio A., Mohr J. J., Wegner G., 2001, ApJ, 555, 558
Rines K., Geller M. J., Diaferio A., Mahdavi A., Mohr J. J., Wegner G., 2002, AJ, 124, 1266
Roche C., et al., 2024, arXiv e-prints, p. arXiv:2402.00928
Rossetti M., et al., 2016, MNRAS, 457, 4515
Rudnick L., 2002, New Astron. Rev., 46, 95
Rudnick L., Owen F. N., 1976, ApJ, 203, L107
Ruiz L. O., Falceta-Gonçalves D., Lanfranchi G. A., Caproni A., 2013, MNRAS, 429, 1437
Sakelliou I., Merrifield M. R., 2000, MNRAS, 311, 649
Schaller M., Robertson A., Massey R., Bower R. G., Eke V. R., 2015, MNRAS, 453, L58
Toro E. F., Spruce M., Speares W., 1994, Shock Waves, 4, 25
Tremblay G. R., et al., 2015, MNRAS, 451, 3768
Vacca V., Murgia M., Govoni F., Feretti L., Giovannini G., Perley R. A., Taylor G. B., 2012, A&A, 540, A38
Vallee J. P., Bridle A. H., Wilson A. S., 1981, ApJ, 250, 66
Ye J.-N., Guo H., Zheng Z., Zehavi I., 2017, ApJ, 841, 45
Yoshida N., Springel V., White S. D. M., Tormen G., 2000, ApJ, 544, L87
Yuan Y., Gu M.-F., Chen Y.-J., 2018, Research in Astronomy and Astrophysics, 18, 108
Zabludoff A. I., Huchra J. P., Geller M. J., 1990, ApJS, 74, 1
MNRAS **000**, 1–21 (2023)

Zabludoff A. I., Geller M. J., Huchra J. P., Ramella M., 1993, AJ, 106, 1301
de Gouveia dal Pino E. M., Benz W., 1993, ApJ, 410, 686

## APPENDIX A: RADIAL AND NON-RADIAL ORBITS FOR THE MODELS H, I AND M

We show in Figure A1 different orbits generated from the variation of $\varphi$ in steps of $60°$ under the assumption of minimum value for $V''_{\text{tot}}$ and considering orbital model H listed in Table 1. We also included $\varphi = 159°$ and $339°$, which are responsible for the radial orbits ($V''_T \cong 0$) seen in Figure A1. Similar plots are displayed in Figure A2 but for the orbital model I that has $\theta = -20°$ against $\theta = 20°$ in the case of model H.

Both figures show that radial orbits do not cross the brighter region of the 3C 338's radio *Ridge*. The same is true for the majority of the values of $\varphi$, reducing substantially the allowed range for this parameter. As $\varphi$ is increased, the whole orbital pattern rotates clockwise in Figure A1 and Figure A2. Given the estimated age of the *Ridge* (∼ 7 Myr), counterclockwise orbits ($\varphi \sim 180°$ and $\sim 0°$ for models H and I, respectively) can be ruled out since they predict a substantial older *Ridge* (e.g., about 60 Myr for $\varphi = 180°$ in the model H).

We show in Figure A3 similar plots for the orbital model Mb that has $\theta = 70°$, compatible with the maximum viewing angle between the normal of the orbital plane of a possible collision between A 2199 and a galaxy group occurred 0.8 Gyr inferred from hydrodynamical *N*-body simulations (Machado et al. 2022). This model does not respect the least orbital velocity condition (see section 2 for further details).

## APPENDIX B: ORBITAL MOTION

The orbital models tested in this work are based on a radio source moving under the influence of gravitational potentials due to the dark matter and the ICM gas in a galaxy cluster. To provide a visualisation of the temporal behaviour of the jet inlet region in our simulations, we present in Figure B1 a sequence of 20 outputs generated every 2 million years from the simulation QEcLt40n13.0v0.1rj1.0, showing the spatial distribution of the temperature in the orbital plane ($z'' = 0$) together with the orbit integrated numerically from the package galpy (see section 2 for further details).

We can note that the jet inlet position varies with time, and specifically in the outputs 18 and 19, the power becomes sufficiently small to cease the plasma ejection in the simulation, with the positions in the orbit coinciding with the gap observed between the ridge and the new jet. An increase in power occurs 1 Myr before the end of this simulation, producing a structure that coincides with the position of the new jet in the output 20.

## APPENDIX C: THE INFLUENCE OF THE INCLINATION BETWEEN JET AND THE ORBITAL PLANE ON ITS PROPAGATION

Some distortions seen in AGN jets may be produced by relative motions of the galaxy and the ICM (e.g., Miley et al. 1972). These tail-like features have been mainly detected in denser environments, such as galaxy clusters (e.g., Rudnick & Owen 1976). The basic idea is that galaxy motion through the ICM produces a ram pressure force that can deform the jet: the force's component perpendicular to the jet bends it towards the opposite direction of the galaxy's velocity, while the parallel component of this force decelerates the jet (e.g., Baan & McKee 1985). This last effect must be behind two results reported in subsection 4.1:

- Jet pointing West are slightly longer than those with a jet oriented easterly independent of the adopted orbital model (Figure 7);
- The increase of the jet viewing angle $\vartheta_{\text{jet}}$ made the projected jet structures more compact (contrary to the usual expectations; see Figure 8).

In both cases, the decrease of the jet length seems to be related to the increase of the $\varepsilon_{\text{jet}}$. To check it is true, we present in Figure C1 the results of five additional simulations adopting orbital model G and evolved for 60 Myr under values for $\varepsilon_{\text{jet}}$ that goes from zero to ninety degrees (jet perpendicular and parallel to the orbital plane, respectively). The jet parameters assumed in those simulations are the same of those in GEcLt60n0.2v0.2rj2.0 (see Table 5 for more details).

We note a larger jet propagation when it is launched perpendicularly to the orbital plane (top right panel in Figure C1), reaching ∼ 91 kpc against ∼ 38 kpc in the parallel case (bottom right panel in Figure C1). To show that a jet launched perpendicularly to the orbital plane does not suffer significant deceleration, we also included in Figure C1 the final (60 Myr) output of a simulation where the jet inlet position remains static and oriented along z-axis ($\varepsilon_{\text{jet}} = 0$) during the whole run. In this case, the jet reached a distance of about 98 kpc, very similar to that found in the simulation adopting $\varepsilon_{\text{jet}} = 0$ and the orbital model G.

Moreover, Figure C1 shows that the orbital motion of the jet inlet region also introduced asymmetries in the number density distribution parallel to the orbital plane ($xy$), stretching it along to the direction where the orbit generated by model G extends. Indeed, all orbital models adopted in this work show this behaviour, suggesting that such feature could be used as an indirect probe for orbital motions of jetted AGNs through the ICM.

This paper has been typeset from a T<sub>E</sub>X/LAT<sub>E</sub>X file prepared by the author.

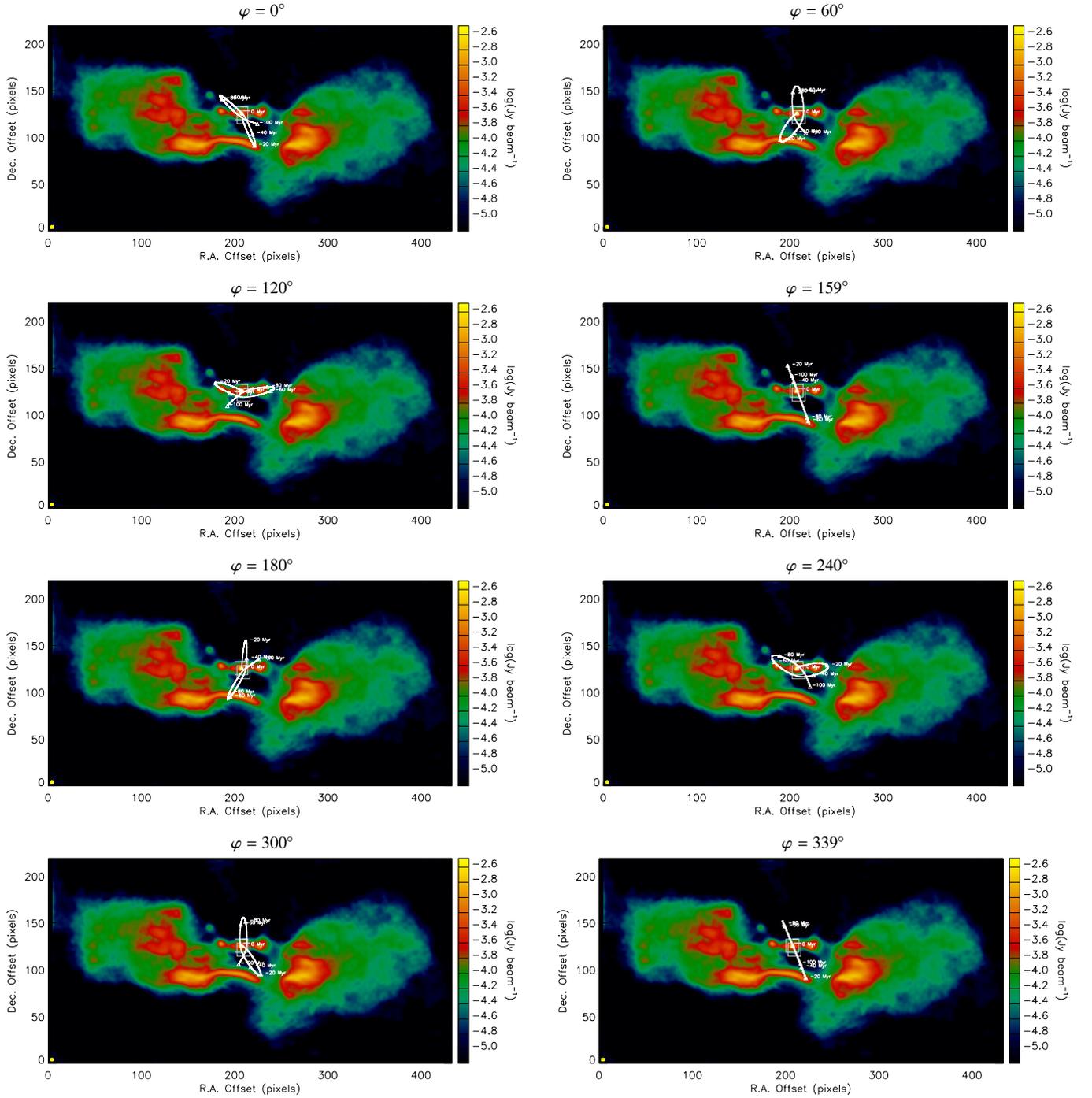

**Figure A1.** Influence of the angle $\varphi$ on the orbits generated by H model under the condition of minimum value for $V''_{\text{tot}}$. Each panel brings a 100-Myr orbit (white lines) for a specific value of $\varphi$ (0°, 60°, 120°, 159°, 180°, 240°, 300° and 339° from upper left to lower right) superposed on the 4.9-GHz image of 3C 338 (1 pixel = 0″.3). The yellow circle at the left bottom of each panel is the CLEAN beam of the radio image.

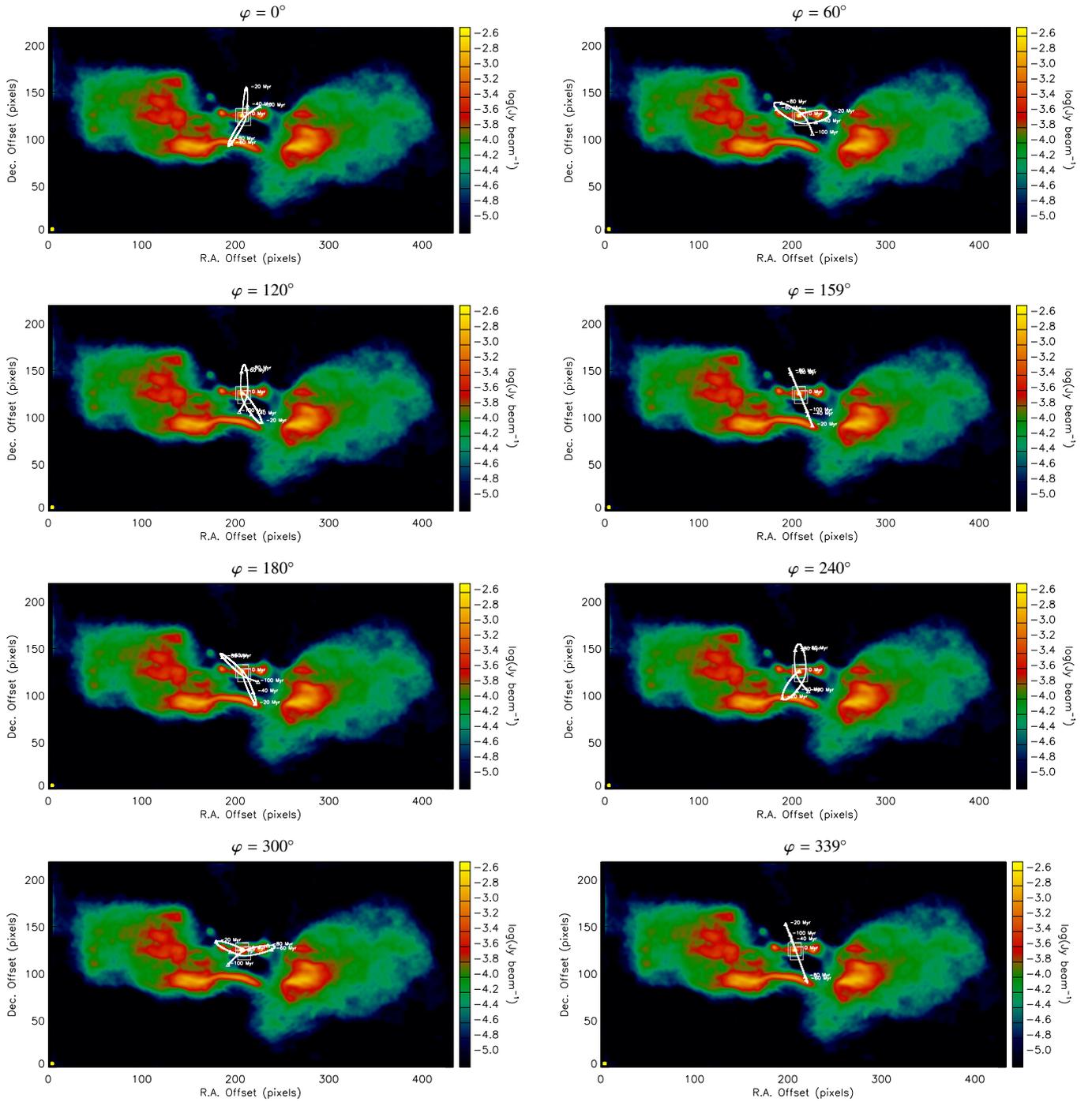

**Figure A2.** The same of Figure A1 but considering orbital model I listed in Table 1.

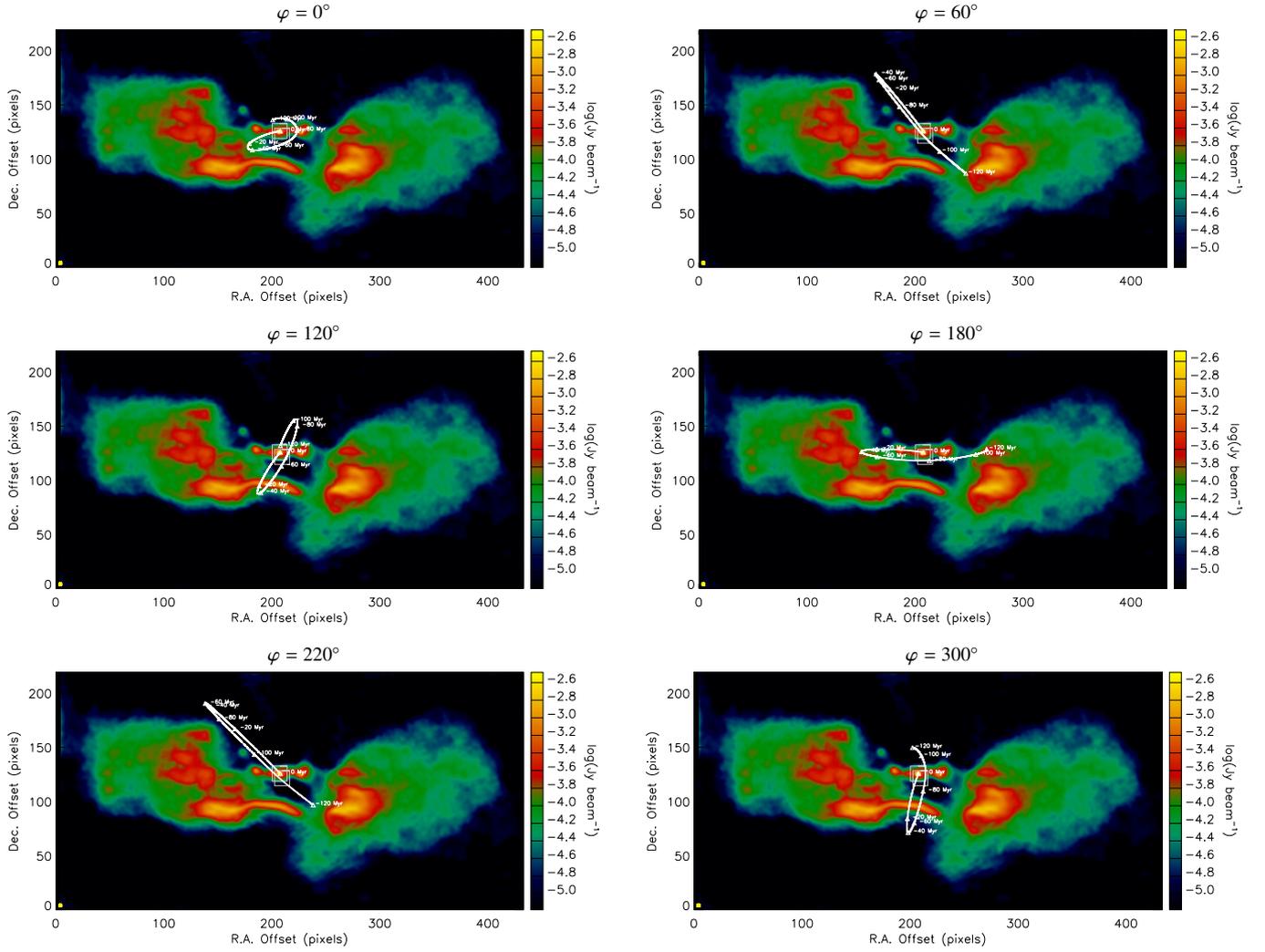

**Figure A3.** Influence of the angle $\varphi$ on the orbits generated by model Mb. Each panel brings a 120-Myr orbit (white lines) for a specific value of $\varphi$ (0°, 60°, 120°, 180°, 220°, and 300° from upper left to lower right) superposed on the 4.9-GHz image of 3C 338 (1 pixel = 0.″3). The yellow circle at the left bottom of each panel is the CLEAN beam of the radio image.

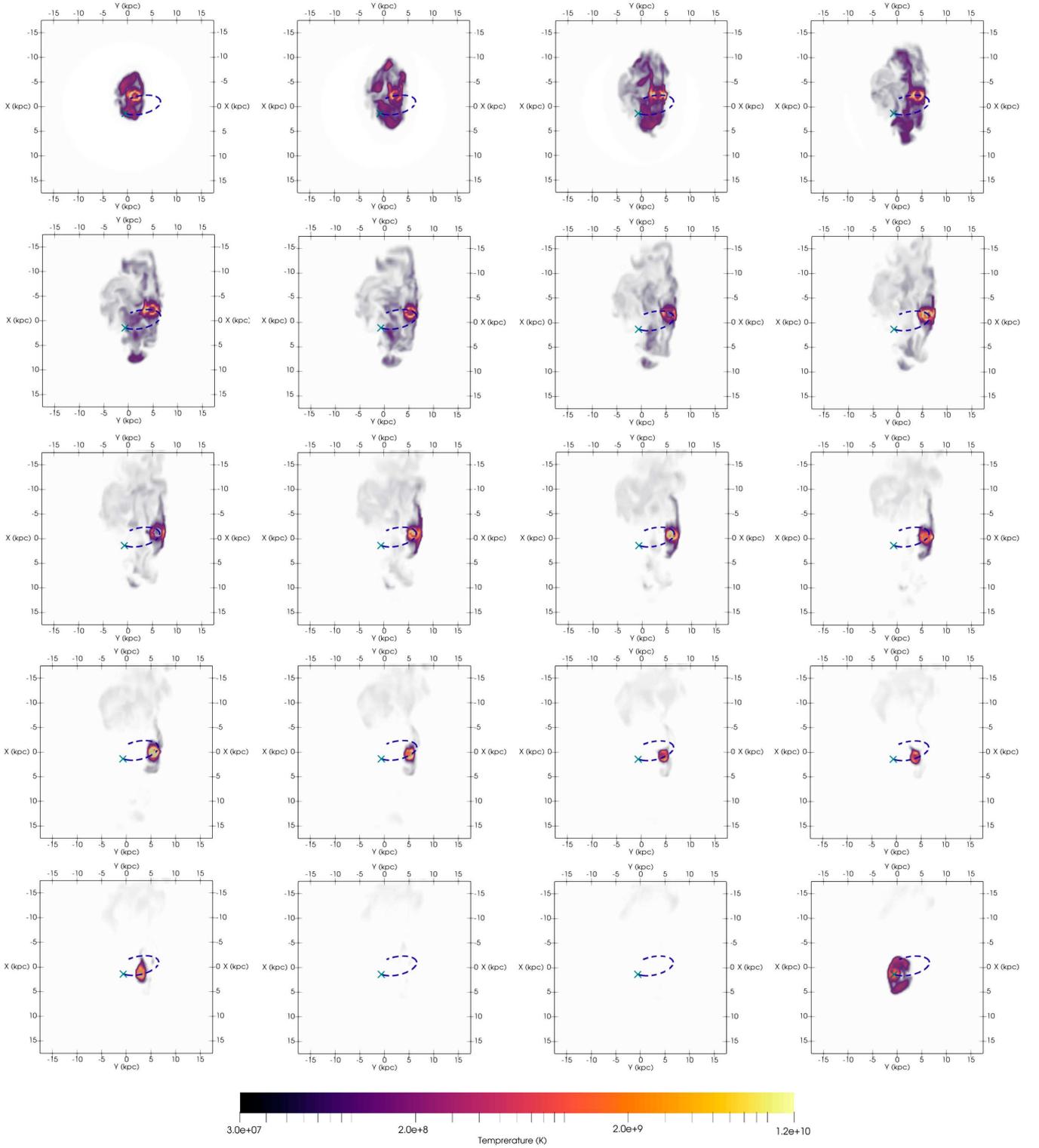

**Figure B1.** Snapshots of the temperature distribution in the orbital plane ($z'' = 0$) generated every 2 Myr from the simulation QEcLt40n13.0v0.1rj1.0. Time evolves from the top left panel to the bottom right one. Blue curves superposed on each panel represent the orbit integrated numerically from the package galpy. The current position of NGC 6166 is marked by the purple cross.

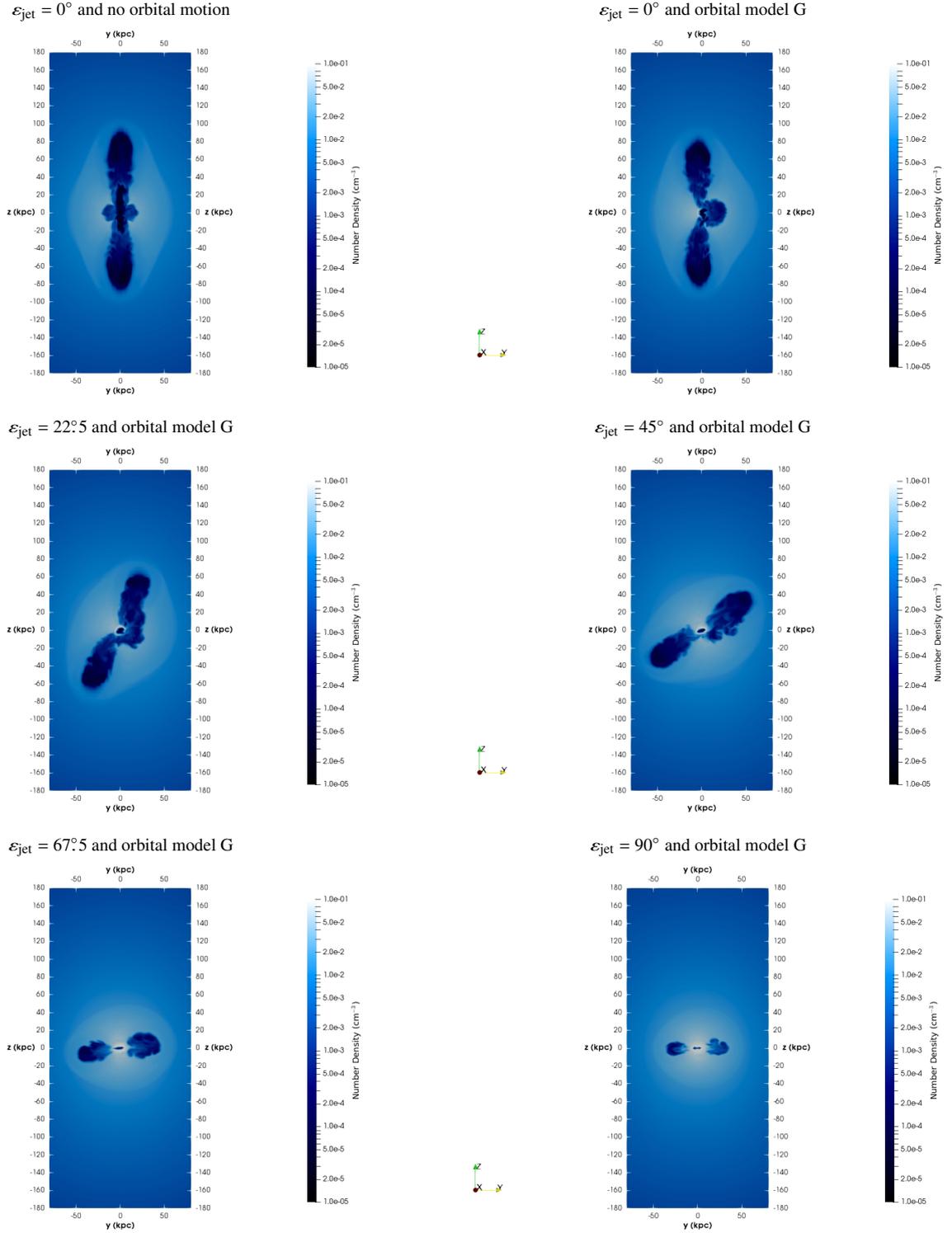

**Figure C1.** Influence of the angle $\varepsilon_\mathrm{jet}$ on the jet propagation after 60 Myr of evolution. The top left panel refers to the case of a static jet inlet region (no orbital motion). The remaining plots show the results from simulations considering a jet inlet region that moves respecting the orbital model G, but adopting five different values for $\varepsilon_\mathrm{jet}$ (from 0° to 90°). Each plot shows the number density distribution at $x'' = 0$.